\documentclass[pre,aps,reprint,noshowpacs,superscriptaddress,floatfix,letterpaper,longbibliography]{revtex4-2}
\usepackage{amsmath,amssymb,amsbsy,amsfonts,amsthm,bbm,bm,mathtools,mathrsfs,bbold}
\usepackage{color}
\usepackage{physics}
\usepackage{xfrac}
\usepackage[dvipsnames]{xcolor}
\usepackage[colorlinks=true,citecolor=MidnightBlue,linkcolor=MidnightBlue,urlcolor=MidnightBlue]{hyperref}
\usepackage{empheq,ragged2e}
\usepackage{pgfplots}
\usepackage{stackengine}
\usepackage{relsize}
\usepackage[inline]{enumitem}
\usepackage[normalem]{ulem}
\usepackage{soul}
\usepackage{comment}
\usepackage{hyperref}
\usepackage{import}
\usepackage{sidecap}
\usepackage[T1]{fontenc}
\usepackage{pgfplots}
\pgfplotsset{compat = newest}
\usetikzlibrary{arrows,intersections}
\usepackage{tikz-3dplot}
\usepackage{tikz}

\usepackage[mathscr]{euscript}
\usepackage{calligra}
\DeclareMathAlphabet{\mathcalligra}{T1}{calligra}{m}{n}
\DeclareFontShape{T1}{calligra}{m}{n}{<->s*[2.2]callig15}{}

\newcommand{\stability}{{\bm{A}}}
\newcommand{\jacobian}{{\bm{M}}}
\newcommand{\ex}{{\bm{x}}}
\newcommand{\bphi}{{\bm{\phi}}}

\newcommand{\yu}{{\bm{u}}}
\newcommand{\by}{{\bm{y}}}
\newcommand{\brho}{{\bm{\varrho}}}

\newcommand{\bx}{{\bm{x}}}

\newcommand{\logder}{\bm{L}}
 
\graphicspath{{../plots/}} %Setting the graphicspath

\usepackage{tcolorbox}
\usepackage{multirow}
\usepackage{graphicx}
\definecolor{new_blue}{RGB}{9, 136, 232}
\newtcolorbox{mybox}[3][]
{%
	colframe = #2!25,
	colback  = #2!10,
	coltitle = #2!20!black,  
	title    = {#3},
	#1,
}
\tcbuselibrary{breakable}
\usepackage{comment}
\usepackage{import}
\usepackage{pgfplots}
\pgfplotsset{compat = newest}
\usetikzlibrary{arrows,intersections}
\usepackage{tikz-3dplot}
\usepackage{tikz}

\begin{document}

\def\xlist{4}
\def\ylist{4}

\title{Phase space contraction rate for classical mixed states}

\author{Mohamed~Sahbani}
\affiliation{Department of Physics,\
	University of Massachusetts Boston,\
	Boston, MA 02125, Boston, USA
}
\author{Swetamber~Das}
\affiliation{Department of Physics,\
	University of Massachusetts Boston,\
	Boston, MA 02125, Boston, USA
}
\affiliation{Department of Physics, School of Engineering and Sciences,\
	SRM University AP, Amaravati, Mangalagiri 522240, India
}
\author{Jason~R.~Green}
\email[]{jason.green@umb.edu}
\affiliation{Department of Chemistry,\
	University of Massachusetts Boston,\
	Boston, MA 02125, Boston, USA
}
\affiliation{Department of Physics,\
	University of Massachusetts Boston,\
	Boston, MA 02125, Boston, USA
}

\date{\today}

\begin{abstract}

Physical systems with non-reciprocal or dissipative forces evolve according to a generalization of Liouville's equation that accounts for the expansion and contraction of phase space volume. 
Here, we connect geometric descriptions of these non-Hamiltonian dynamics to a recently established classical density matrix theory. 
In this theory, the evolution of a ``maximally mixed'' classical density matrix is related to the well-known phase space contraction rate that, when ensemble averaged, is the rate of entropy exchange with the surroundings. 
Here, we extend the definition of mixed states to include statistical and mechanical components, describing both the deformations of local phase space regions and the evolution of ensembles within them.
As a result, the equation of motion for this mixed state represents the rate of contraction for an ensemble of dissipative trajectories. 
Recognizing this density matrix as a covariance matrix, its contraction rate is another measure of entropy flow characterizing nonequilibrium steady states.

\end{abstract}

\maketitle

\section{Introduction}

Classical statistical mechanics is based on Liouville's theorem and equation~\cite{zwanzig2001nonequilibrium}.
Perhaps the best-known statements of Liouville's equation are for Hamiltonian systems~\cite{aubin2012nonlinear, abraham2012manifolds} in which it is a conservation law for phase space volumes, probability densities $d_t\ln\rho = 0$, and energy.
However, when dynamical systems are dissipative, driven, stressed or constrained~\cite{evansStatisticalMechanicsNonequilibrium}, they evolve in a phase space that is compressible. 
To describe these non-Hamiltonian dynamics, the usual statements of Liouville's equation and theorem ``generalize'' to explicitly include the rate of phase space compressibility $\Lambda = -d_t\ln\rho$~\cite{gerlichVerallgemeinerteLiouvilleGleichung1973,prugovevcki1978liouville} -- the rate at which phase space expands and contracts in response to flows of energy in and out of the system~\cite{dorfmanIntroductionChaosNonequilibrium1999,gaspardChaosScatteringStatistical1998,evansStatisticalMechanicsNonequilibrium}. 
This generalization is useful in designing molecular dynamics algorithms for nonmicrocanonical ensembles and stationary nonequilibrium flows~\cite{tuckerman2001non,ezra2004statistical} but also defining nonequilibrium observables. 
The phase space compressibility or ``contraction rate'' is related to the entropy production/flow~\cite{Dorfman2020} and, for many-particle systems interacting with Gaussian and Nos\'e-Hoover thermostats, the electrical conductivity, diffusion, and viscosity~\cite{gaspardTransportPropertiesLyapunov1990,dorfmanChaoticScatteringTheory1995a,Dorfman1997-sy}.
As averages of $\Lambda$ over the phase space distribution, these statistical mechanical observables require both $\Lambda$ and $\rho$ over time.

The statistical form of the contraction rate defined through the generalized Liouville's equation has a complementary mechanical form that derives from Jacobi's formula. Compressible flows generate coordinate transformations in phase space with a Jacobian matrix $\jacobian$. Its determinant evolves in time according to \textit{Jacobi's formula}~\cite{muir_treatise_determinants_1933}:
\begin{equation}
	\label{eq:jacobi1}
	\Lambda := \frac{d}{dt} \ln  \lvert\boldsymbol{M}\rvert = \Tr\left(\boldsymbol{M}^{-1}\frac{d\boldsymbol{M}}{dt}\right),
\end{equation}
{which we can see as another, purely mechanical, expression of the contraction rate in} $\Lambda = -d_t\ln\rho$~\cite{Andrey1985a,Andrey1985b,Ramshaw1986remarks}.
Hamiltonian systems are clearly a special case in which incompressible flows $\Lambda = 0$ make both $\lvert\boldsymbol{M}\rvert$ and $\rho$ constants of motion.
This version of Jacobi's formula is a part of treatments of phase space as a Riemannian manifold~\cite{betancourt2013general}.
For example, Tuckermann, Mundy, and Martyna showed that the determinant of the metric describing the Riemannian geometry evolves at a rate determined by the phase space compressibility $\Lambda$~\cite{tuckerman1999classical}.
Ramshaw~\cite{Ramshaw_2002} later showed that a coordinate transformation of the phase space volume element makes $\Lambda = -d_t\ln\rho$ covariant, establishing an equivalence between the known generalizations of Liouville's equation.

Recently, two of us established an alternative generalization of Liouville's equation using a classical density matrix~\cite{DasGreen2022}. 
This density matrix consists of perturbation (tangent) vectors of classical systems, which grounds it in dynamical systems theory. 
While it is classical, this theory is analogous to the density matrix formulation of quantum mechanics based on wavevectors in Hilbert space.
It includes uncertainty relations and speed limits~\cite{DasGreen2023speed,DasGreen2024} using a Fisher information for (local) phase space geometry~\cite{Sahbani2023classical}. 
These features derive from the classical density matrix, which is similar to the metric tensor of the underlying phase space and has a determinant that satisfies Liouville's theorem and equation~\cite{DasGreen2022}. 
In this theory, $\Lambda$ preserves phase space volumes in non-Hamiltonian dynamics of the density matrix.

\begin{figure*}[t]
	\includegraphics[width=0.9\textwidth]{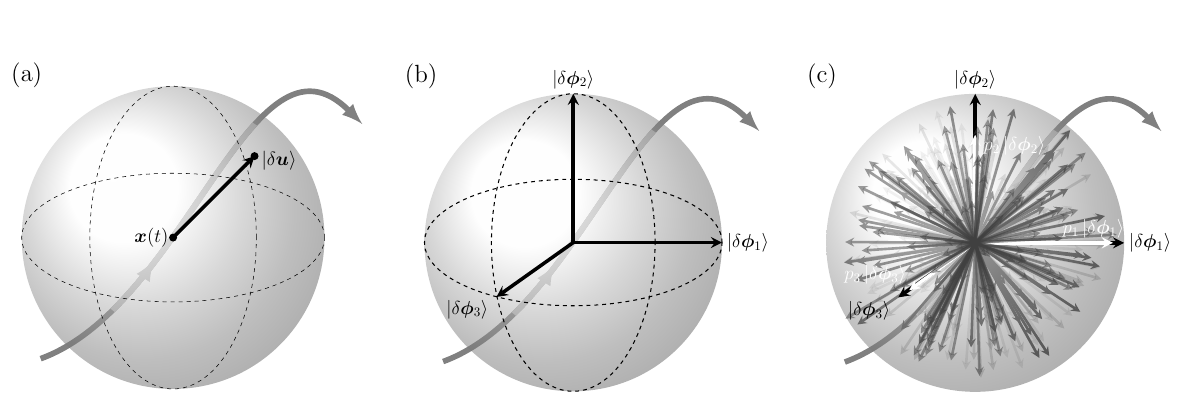}
	\caption{\label{fig:mixed}\textit{Illustration of classical density matrix states at a time $t$ in a volume of phase space $\delta V$ comoving with a point $\boldsymbol{x}$ along a trajectory.} (a) Pure states $\brho_p=\dyad{\delta\boldsymbol{u}}$ consist of a single unit tangent vector $\ket{\delta\boldsymbol{u}}$ (black). (b) Maximally mixed states $\brho_{\textrm{max}}=d^{-1}\sum_i^d\dyad{\delta\bphi_i}$ consist of a set of $d$ basis vectors $\ket{\delta\bphi_i}$ (black) with probability $1/d$. Here, $d=3$. (c) Fully mixed states $\brho=\sum_i^dp_i\dyad{\delta\bphi_i}$ represent an ensemble of tangent vectors in the $d$-dimensional eigenbasis with probabilities $p_i$.}
\end{figure*}

These two independent lines of work on non-Hamiltonian systems raise questions that we investigate here. 
First, what is the relationship between this density matrix theory and the previous work on the generalized Liouville equation ~\cite{sergi2003non, ezra2004statistical,sergi2007geometry, bravetti2015liouville}? 
The second question comes from the fact that the density matrix theory has so far largely focused on ``pure'' states, Fig.~\ref{fig:mixed}(a).
For many-particle systems, pure classical density matrices entirely mechanical -- being composed of perturbations to phase space variables. 
However, fully mixed states would consist of statistical mixtures, ensembles of classical systems, together with perturbations capturing the structure of phase space, Fig.~\ref{fig:mixed}(b,c). 
What is the phase space contraction rate $\Lambda$ and Jacobi's formula for mixed states and how is this rate related to entropy rates?
While statistical-mechanical observables generally require the dynamics of both $\Lambda$ and $\rho$, we show that encoding their joint dynamics in a classical density matrix leads to an ensemble-level contraction rate and a local measure of entropy flow.

In Section~\ref{sec:cmd_theory}, we review the necessary classical density matrix theory and define mixed states.
The derivation of Jacobi's formula for pure states in Section~\ref{sec:3} establishes the connection between the density matrix theory and previous geometric descriptions of Liouville's equation.
In Section~\ref{sec:4}, we discuss the special case of maximally mixed states, connecting them to Jacobi's formula in Eq.~\ref{eq:jacobi1} and defining the corresponding phase space contraction rate.
We analyze a classical damped harmonic oscillator as an example.
In Section~\ref{sec:5}, we extend the notion of ``state'' from pure to mixed states and derive a generalization of Liouville's equation that accounts for the evolution of statistical ensembles
with a classical density matrix.
We obtain a generalization of Jacobi's formula for the phase space compressibility of an ensemble of infinitesimal perturbations in Section~\ref{sec:6}.
This Jacobi formula for mixed states defines a new rate for the compressibility of ensembles of phase points.
While the rates we define refer to expansion and contraction, we will often have dissipative systems in mind and refer to them simply as ``contraction rates''.
In Section~\ref{sec:7}, we directly relate this contraction rate to the Gibbs entropy rate for classical statistical mixtures.

\section{Classical density matrix theory} \label{sec:cmd_theory}

Before deriving the phase space contraction rates and Jacobi's formulas for each density matrix, let us first describe the ingredients of the classical density matrix theory that we will need. 

Consider a classical trajectory $\ex(t) := [x^1(t), x^2(t), \ldots , x^n(t)]^\top$ in the $n$-dimensional state space of a differentiable dynamical system $\dot{\ex} = \boldsymbol{F}(\ex)$. 
An infinitesimal perturbation vector on the trajectory represented by a finite-dimensional column vector with the ket $\ket{\delta\ex(t)} := [\delta{x^1(t)}, \delta{x^2}(t), \ldots , \delta{x^n(t)}]^{\top}$ in a real vector space points from the reference trajectory to a nearby phase point. 
Such a vector evolves according to the linearized equation of motion
\begin{align}\label{eq:linear-transformation} 
	\frac{d}{dt} \ket{\delta\ex(t)} = \boldsymbol{\stability}[\ex(t)] \ket{\delta{\ex(t)}},
\end{align}
where the elements of the stability matrix $\stability: = \stability[\ex(t)] = \boldsymbol{\nabla} \cdot \boldsymbol{F}$ are $(\stability)^i_j = \partial \dot x^i/\partial x^j$. 
This perturbation vector $\ket{\delta\ex(t)}$ may stretch or contract and rotate as it comoves with the trajectory. 
For example, on a chaotic trajectory, the norm of $\ket{\delta\ex(t)}$ grows exponentially with time and must be periodically normalized in order to compute the rates of stretching/contracting, Lyapunov exponents~\cite{pikovskyLyapunovExponentsTool2016}. 
A unit perturbation vector $\ket{\delta{\yu(t)}}=\ket{\delta{\ex(t)}}/\langle \delta \ex|\delta \ex\rangle^{1/2}$ has the dynamics 
\begin{align}\label{eq:EOM-unit} 
	\frac{d}{dt}\ket{\delta \yu(t)} =  \bar\stability\ket{\delta \yu(t)}.
\end{align}
Here, $\langle \delta \ex|\delta \ex\rangle^{1/2}$ represents the $\ell^2$ norm, the matrix $\bar\stability = \stability - \bra{\delta \yu}\stability\ket{\delta\yu}\boldsymbol{\mathbb{1}}$ measures the deviation from the matrix $\stability$, and $\boldsymbol{\mathbb{1}}$ indicates an $n\times n$ identity matrix.
The matrix $\bar\stability$ guarantees that the evolution of $\ket{\delta \yu(t)}$ over time preserves the norm, $\bra{\delta\yu(t_0)}\ket{\delta \yu(t_0)} = \bra{\delta\yu(t)}\ket{\delta \yu(t)}$. 

Analogous to quantum mechanics~\cite{Chenu2024}, we can define an alternative state for these dynamics. 
Perhaps the most fundamental definition of the classical density matrix is as a projection operator $\brho_p$ using the outer product
\begin{equation}
	\brho_p(t) = \dyad{\delta \yu(t)}{\delta \yu(t)}\quad\quad\text{(pure)}.
\end{equation}
This definition ensures $\brho_p$ is positive semidefinite with $\Tr\brho_p = 1$ and $\brho_p^2=\brho_p$. 
Pure quantum states have similar properties~\cite{shankar2012principles,sidhuGeometricPerspectiveQuantum2020}, so we refer to $\brho_p$ as a \textit{pure perturbation state}. 
Using Eq.~\ref{eq:EOM-unit}, its equation of motion is
\begin{align}\label{eq:EOM-pure}
  \frac{d}{dt}\brho_p = \bar\stability \brho_p + \brho_p\bar\stability^\top.
\end{align}
These dynamics preserve the trace of $\brho_p$ at all times because the trace of the right side vanishes $\Tr (\bar\stability\brho_p) = \Tr(\bar\stability^\top\brho_p)$.
The solution of Eq.~\ref{eq:EOM-pure} 
evolves the normalized state $\brho_p$ from the initial time $t_0$ to the time $t$:
\begin{align}\label{eq:evol-eq3}
    \brho_p(t) = \widetilde\jacobian(t,t_0)\brho_{p}(t_0)\widetilde\jacobian^\top(t,t_0).
\end{align}
%\begin{align}\label{eq:evol-eq3}
 %   \ket{\delta{\yu}(t)} =  \widetilde{\jacobian}(t,t_0)\ket{\delta{\yu}(t_0)}.
%\end{align}
The propagator matrix $\widetilde{\jacobian}(t,t_0)$ is the solution (App.~\ref{SM:E1}) to
\begin{align}\label{eq:jc4}
 \frac{d\widetilde{\jacobian}}{dt} = \bar{\stability}\widetilde{\jacobian}.
\end{align}
More explicitly, the propagator
\begin{align}\label{eq:jc5}
     \widetilde{\jacobian}(t,t_0) = \mathcal{T}e^{\int_{t_0}^{t}{\bar{\stability}}(t')dt'}
\end{align}
includes time ordering $\mathcal{T}$ because $\bar\stability$ depends on time $t$ and the stability matrices $\bar\stability(t_i)$ and $\bar\stability(t_j)$ at two different times $t_i$ and $t_j$ do not generally commute: $[\bar\stability(t_i),\bar\stability(t_j)]\neq 0$.
This propagator has several distinctive properties (App.~\ref{SM:E2}): it is initially $\widetilde{\jacobian}(t_0,t_0) = \boldsymbol{\mathbb{1}}$ the identity matrix; $\widetilde{\jacobian}^{-1}(t,t_0) =\widetilde{\jacobian}(t_0,t)$; the composition property $\widetilde{\jacobian}(t_2,t_0) =\widetilde{\jacobian}(t_2,t_1)\widetilde{\jacobian}(t_1,t_0)$; and the time reversal property $\widetilde{\jacobian}(t,t_0)\widetilde{\jacobian}(t_0,t) = \boldsymbol{\mathbb{1}}$.

Other aspects of this theory in Refs.~\cite{DasGreen2022,Sahbani2023classical,DasGreen_Spectralbounds} include a classical Fisher information and speed limits on dynamical observables. However, these features of the theory are unnecessary for the present discussion. 

%\jrg{\sout{is the norm preserving operator~\cite{fernandez1989time} \jrgc{Why is this reference put here?} which}} 

%So  $\bar\stability(t_i)$ and  $\bar\stability(t_j)$~\cite{lam1998decomposition}.  
%So $\boldsymbol{\stability}(t_i)$ and $\boldsymbol{\stability}(t_j)$ don't commute in general, $[\boldsymbol{\stability}(t_i),\boldsymbol{\stability}(t_i)]\neq{0}$.

The focus of this theory so far has been ``pure'' perturbation states, so we will introduce a more general classical density matrix that is a statistical mixture of weighted pure states. 
This density matrix represents an ensemble of phase space points in the neighborhood of a point on the reference trajectory, Fig~\ref{fig:mixed}.
%\ms{ \sout{ Like the interpretation of the single normalized perturbation state $\ket{\delta\yu(t)}$, the mixed state can be visualized as many  random disturbances in any direction at a specific point of the phase space trajectory, causing significant deviations from the intended path. These disturbances have more effect than the case of the pure state,   affecting also significantly the compressibility of the trajectory.} }
We can express these \textit{mixed perturbation states},
\begin{align}\label{eq:8}    
	\brho := \brho(\ex,t) = \frac{1}{k} \sum_{j}^{k}\dyad{\delta \yu_{j}}{\delta \yu_{j}}   
	= \sum_i^d p_i\dyad{\delta \bphi_{i}}{\delta \bphi_{i}}
\end{align}
%\begin{equation}\label{eq:8}
	%\brho =  \sum_{i=1}^d p_i\brho_i\quad\quad\text{(mixed)},
%\end{equation}
in terms of the $d$ eigenvectors $\ket{\delta \bphi_i}$ and eigenvalues $p_i$ of $\brho$ with $\sum_{i=1}^d p_i = 1$.
These states have both statistical and mechanical contributions that give them a purity between the extremes $1/d\leq \Tr(\brho^2)\leq1$.
The equation of motion is the Lyapunov equation, $d_t\brho = \left(\logder\brho + \brho \logder^{\top}\right)/2$, implicitly defining the logarithmic derivative $\logder$. 

The logarithmic derivative $\logder$ is (App.~\ref{SM:Logarithmic-derivative})
\begin{align}\label{eq: Logarithmic-derivative}
\begin{split}
    \boldsymbol{L} &= 2\sum_{i,j=1}^{d} \frac{\bra{\delta\bphi_i}d_t\brho\ket{\delta\bphi_j}}{p_{i} + p_{j}}\dyad{\delta\bphi_i}{\delta\bphi_j}\\
    &+ \sum_{i,j=1}^{d}\frac{2p_{i} }{p_{i} + p_{j}}\bra{\delta\bphi_i} \boldsymbol{L}_{-}\ket{\delta\bphi_j} \dyad{\delta\bphi_i}{\delta\bphi_j},
\end{split}
\end{align}
which has both symmetric $\boldsymbol{L}_+$ and anti-symmetric parts $\boldsymbol{L}_{-}$. This expression also includes the eigenvalues $\{p_i\}$ and eigenvectors $\{\ket{\bphi_i}\}$ of the density matrix $\brho$ in the $d$-dimensional eigenspace. 
Here, if $\boldsymbol{L} = \boldsymbol{L}_+$ is symmetric Eq.~\ref{eq: Logarithmic-derivative} reduces to (App.~\ref{SM: Symmetric-Logarithmic-derivative})
\begin{align}\label{eq: Symmetric-Logarithmic-derivative}  
\begin{split}
	\boldsymbol{L}=\boldsymbol{L}_+ &= \sum_{k=1}^{d} \frac{d\ln p_k}{dt} \dyad{\delta\bphi_k} \\ &+ 2\sum_{i\neq{j}}^{d}\frac{\left(p_{j}-p_{i}\right)}{p_{j} + p_{i}}\bra{\delta\bphi_i}\ket{d_t{\delta\bphi_j}} \dyad{\delta\bphi_i}{\delta\bphi_j}.
\end{split}
\end{align}
While the classical logarithmic derivative is rarely symmetric~\cite{DasGreen2023speed}, in that case $\boldsymbol{L}$ does become the classical analog of the quantum Fisher information~\cite{Helstrom1967,sidhuGeometricPerspectiveQuantum2020}.

%\jrgc{Does Eqs. 5-7 hold for mixed states too? We seem to avoid answering this question.} \msc{They don't hold as we discussed.}

Pure states are one special case of Eq.~\ref{eq:8} with $p_i=1$ and $p_j=0$ for all $j\neq i$ and $\ket{\delta \bphi_i} =\ket{ \delta \yu_{i}}$.
Since they consist of a single vector, they are the simplest states, are entirely mechanical, and have maximum purity $\Tr(\brho_i^2)=1$. 
Another special case are maximally mixed states
\begin{equation}\label{eq:9}
	\brho_{\text{max}} %= k^{-1}\sum_{\ms{j}=1}^{\ms{k}} \brho_{\ms{j}} 
	= d^{-1}\sum_{i=1}^d \dyad{\delta \bphi_i} \quad\quad\text{(maximally mixed)}
\end{equation}
with a complete a set of equally weighted basis states, $p_i=1/d$. Since the density matrix is symmetric, we have the completeness relation  $\sum_{i=1}^d \dyad{\delta \bphi_i}=\boldsymbol{\mathbb{1}}$. 
Completely mixed states are also purely mechanical with each pure state contributing equally and minimum purity, $\Tr(\brho_\text{max}^2)=1/d$. 
%$\Tr(\brho_\text{max}^2) = \Tr(d^{-1}\boldsymbol{\mathbb{1}}d^{-1}\boldsymbol{\mathbb{1}})=d^{-2}\Tr\boldsymbol{\mathbb{1}}=d^{-2}d=d^{-1}$}.
Hence, $\brho_{\text{max}} = d^{-1}\boldsymbol{\mathbb{1}}$ and $\boldsymbol{L}=0$. Given the equation of motion $d_t\brho_\text{max} =  \left(\logder\brho_\text{max} + \brho_\text{max}\logder^\top\right)/2 =0$, we see that these states are conserved over time.

The density matrix and its equation of motion are the ingredients we need to establish Liouville's theorem/equation and new phase space contraction rates through Jacobi's formula.

\section{Jacobi's formula for pure states}\label{sec:3}

Jacobi's formula is a more general statement that any arbitrary matrix $\boldsymbol{B}$ satisfies $\frac{d}{dt} |\boldsymbol{B}(t)|=|\boldsymbol{B}(t)| \Tr\left(\boldsymbol{B}(t)^{-1} \frac{d \boldsymbol{B}(t)}{d t}\right)$.
So, in the present context, we can derive Jacobi's formula and define ``contraction rates'' for the propagator $\widetilde{\jacobian}$ as well as the density matrices $\brho$.
The former complement the known Jacobi formula $\Lambda = d_t\ln |\boldsymbol{M}|$ in Eq.~\ref{eq:jacobi1}, and the latter complement the generalized Liouville equation $\Lambda=-d_t\ln\rho$.

From the pure state definition, we can see that the determinant is time invariant $d_t\lvert \brho_p \rvert = 0$, akin to Liouville's equation.  
The density matrix $\brho_p$ is not invertible because the determinant of the density matrix is zero and $\Tr(\brho_p^{-1} d_t\brho_p)$ is undefined. Consequently, pure states do not satisfy a Jacobi formula.
%\begin{align}\label{eq:jcX}
%	\frac{d\ln\lvert \brho_p \rvert}{dt} = \Tr(\brho_p^{-1} \frac{d\brho_p}{dt}) = 0,
%\end{align}
%because $\brho_p$ is singular. 
%Meanwhile, the zero-determinant $\lvert \brho_p \rvert=0$ satisfies the equation of motion \sdas{trivially}:
%\begin{align}\label{eq:jcX}
%	\frac{d\lvert \brho_p \rvert}{dt} = 0, 
%\end{align}
However, we can derive a contraction rate through Jacobi's formula with $\widetilde{\jacobian}$. 

To derive this formula for the propagator $\widetilde\jacobian$, we start with the equation of motion for its determinant.
Using the trace linearity and the relations  $\lvert{e}^{\boldsymbol{B}}\lvert = e^{\Tr\left(\boldsymbol{B}\right)}$ and $\lvert\boldsymbol{B}\boldsymbol{C} \rvert =  \lvert\boldsymbol{B}\rvert \lvert\boldsymbol{C}\rvert$ for matrices $\boldsymbol{B}$ and $\boldsymbol{C}$~\cite{olver2006applied}, we obtain the determinant (App.~\ref{SM:E3}):
%\begin{align*}
%\begin{split}
%     \lvert \boldsymbol{M} \rvert &= \det{\boldsymbol{M}(t,t_0)} \\
%     &=\lvert\mathcal{T}e^{\int_{t_0}^{t}\boldsymbol{\bar{\stability}}(t')dt'}\rvert\\
%     &= \lvert{e^{\int_{t_0}^{t}\boldsymbol{\bar{\stability}}(t')dt'}}\rvert\\
%     &= e^{\Tr\left(\int_{t_0}^{t}\boldsymbol{\bar{\stability}}(t')dt'\right)}\\
%     &=  e^{\int_{t_0}^{t}\Tr\left(\boldsymbol{\bar{\stability}}(t')\right)dt'}
%\end{split}
%\end{align*}
\begin{align}\label{eq:jc6}
    {\lvert \widetilde\jacobian \rvert} = e^{\int_{t_0}^{t}\Tr({\bar{\stability}}(t'))dt'},
\end{align}
where $|\cdot|$ represents the determinant and we suppress the time dependence $\widetilde\jacobian(t,t_0)$. 
Its time derivative
\begin{align}\label{eq:jc7}
    \frac{d{\lvert  \widetilde \jacobian \rvert}}{dt} =\lvert  \widetilde \jacobian \rvert \Tr({\bar{\stability}})
\end{align}
gives the rate of change of the determinant of the norm preserving propagator $\widetilde \jacobian$.
Using Eq.~\ref{eq:jc4}, we find Jacobi's formula for $\widetilde\jacobian$ (App.~\ref{SM:E4}):
\begin{equation}
\begin{aligned}\label{eq:jc8}
	\Lambda_{p} &:= \frac{d}{dt}\ln\lvert \widetilde\jacobian \rvert
				= \Tr(\widetilde\jacobian^{-1} \frac{d\widetilde\jacobian}{dt})
	%\lvert \widetilde\jacobian \rvert 
				= \Tr(\bar\stability)\\
				&= \Lambda-d\langle\stability\rangle
				= \frac{1}{2}\Tr \logder_{p}
\end{aligned}
\end{equation}
and define the contraction rate for pure states $\Lambda_p$. 
This Jacobi formula indicates how quickly the volume scaled by the propagator $\widetilde\jacobian$ matrix changes over time. 
For comparison, this Jacobi formula has the same form as that used in previous work on compressible flows: $\Lambda = d_t \ln |\boldsymbol{M}| =  \Tr\left(\boldsymbol{M}^{-1}d_t\boldsymbol{M}\right)$~\cite{magnus2019matrix}. 
%Here, the pure classical density matrix uses the perturbation of the phase-space flow. 
From Eq.~\ref{eq:jc8}, we can also see that the two contraction rates they define are linearly related. 
The difference is whether the contraction rate derives from the compressibility of the phase space defined by the coordinate transformation $\jacobian$ or its norm-preserving counterpart $\widetilde{\jacobian}$. 

The second line in Eq.~\ref{eq:jc8} expresses the pure state contraction rate in terms of the logarithmic derivative $\logder_{p} = 2\bar\stability$ for the pure state in $d$ phase space dimensions $d$.
The trace of $\logder_{p}$ measures the deviations between the well-known phase space contraction rate and the instantaneous Lyapunov exponent, 
$\Tr({\bar{\stability}}) = \Tr(\stability)-\Tr(\stability \brho_p\boldsymbol{\mathbb{1}}) = \Lambda-d\langle\stability\rangle$~\cite{DasGreen2022}.  
As we will see, the other density matrices have identical expressions for their Jacobi formulas and ``contraction rates'' in terms of the logarithmic derivative.

\subsection{Example: Damped harmonic oscillator}

As an example, we consider the one-dimensional linear damped harmonic oscillator~\cite{goldstein2002classical} with mass $m$ and frequency $\omega$. Its equations of motion are: 
\begin{align}\label{eq:Damped-equations-of-motion}
\dot{q}= \frac{p}{m}\qquad          \dot{p}=-m{\omega^2}q - \frac{\gamma}{m}p,
\end{align}
where $\gamma$ is the positive damping coefficient. The stability matrix of these dissipative dynamics is
\begin{align}\label{eq:damped-stability-matrix}
    \stability =  \begin{pmatrix} 
	0&1/ma\\
	-ma\omega^2&-\gamma/m\end{pmatrix}.
\end{align}
The parameter $a$ has dimensions $[\text{TM}^{-1}]$, which ensures the stability matrix elements have a dimension of $[\text{T}^{-1}]$~\cite{Sahbani2023classical}. From this stability matrix, the phase space contraction rate for the damped harmonic oscillator is $\Lambda =\Tr\stability=-\gamma/m$. This expression indicates the rate at which the system loses energy as trajectories decay to the fixed point. 
The pure state contraction rate is then $\Lambda_p =  -\gamma/m - 2\langle\stability\rangle$. Physically, it measures the deviation between the instantaneous Lyapunov exponent $-2\langle\stability\rangle$ scaled by $d=2$ and the phase space contraction rate $\Lambda$.

To find $\langle\stability\rangle = \bra{\delta \yu}\stability\ket{\delta \yu}=\Tr(\stability\brho_p)$ over a pure state $\brho_p(t) = \dyad{\delta \yu(t)}{\delta \yu(t)}$, we use %$\ket{\delta \yu}$ as the normalized phase space velocity} $(\dot q, \dot p)^\top$ %leads to    $\ket{\delta \yu} = \bra{\delta \ex}\ket{\delta \ex}^{-1/2} \ket{\delta \ex}$, which can be written explicitly in the $2$-dimensional phase space:
\begin{align}\label{eq:perturbed-pure-state}
     \ket{\delta\yu}=\frac{({\dot{q}}, {\dot{p}})^\top}{\sqrt{{\dot{q}^2} + {\dot{p}^2}}} =\begin{pmatrix}
	u\\
	v\end{pmatrix},
\end{align}
which satisfies the normalization condition $\bra{\delta \yu}\ket{\delta \yu} = u^2+v^2 = 1$ at all times. The expectation value of $\stability$
\begin{align}
\langle\stability\rangle &= \frac{v}{ma}\left[\left(1-\left(ma\omega\right)^2\right)u - \gamma{av} \right]
\end{align}
leads to the pure state contraction rate
%This expression of the expectation value can be simplified by using the fact $ v^2 = 1-u^2$ and plugging in it in the expression of $\Lambda_p$. Hence, we get   
\begin{align}
 \Lambda_p &=  2\frac{ \left(ma\omega\right)^2-1}{ma}uv + \frac{\gamma}{m}\left(v^2 -u^2\right),
\end{align}
which decays along trajectory until $u$ and $v$ vanish.

For the simple harmonic oscillator, there is no damping coefficient, $\gamma =0$, and the energy of the system is conserved. Phase space orbits trace an ellipse because of the incompressible phase flow, and the usual contraction rate is $\Lambda=0$. As a result, the pure state contraction rate reduces to
\begin{align}
 \Lambda_p &=  2\frac{ \left(ma\omega\right)^2-1}{ma}uv.
\end{align}
This rate oscillates because the normalized phase space coordinates $u$ and $v$ contain the periodic solutions of Eq.~\ref{eq:Damped-equations-of-motion}.
In this case, the propagator $\widetilde\jacobian$ transports the (normalized) phase space velocity vector $\ket{\delta \yu(t_0)}$ around an elliptical trajectory with a constant magnitude.  The rate $\Lambda_p$ vanishes only when perturbations in configuration $u=\dot q/\|(\dot q, \dot p)^\top\|$ and momentum $v=\dot p/ \|(\dot q, \dot p)^\top\|$ are zero.

So far, we have discussed the simplest case of a single tangent vector. We now extend the phase space contraction rate to a mixture of these pure states. 

\section{Jacobi's formula for maximally mixed states}\label{sec:4}

Now that we have Jacobi's formula for pure states, we can derive Jacobi's formula for the maximally mixed states of non-Hamiltonian dynamics. 
As in Eq.~\ref{eq:9}, these states consist of a set of vectors that span the phase space. 
To span a phase space volume in $d$-dimensions, we need a complete set of $d$ basis vectors.
The propagator $\widetilde \jacobian$ evolves each from a time $t_0$ to a time $t$. 
We will call the density matrix for the $i^\text{th}$ vector, $\brho_i = \dyad{\delta \yu_i}{\delta \yu_i}$, the $i^\text{th}$ basis state.
Each $\brho_i$ evolves according to Eq.~\ref{eq:EOM-pure} and a set of $\widetilde{\boldsymbol{M}}$ evolves according to Eq.~\ref{eq:jc7}.
%We will show that these exponents restore the preservation of the volume.
For a $d$ dimensional state space, we rewrite Eq.~\ref{eq:jc8} as
\begin{align}\label{Eq:det-pure}
	\Lambda_{p,i} = \frac{d}{dt}\ln|\widetilde{\jacobian}_i| = \Tr \bar{\stability_i} = \frac{1}{2}\Tr \logder_{p,i},
\end{align}
in terms of the logarithmic derivative $\logder_{p,i}$ for the $i$th pure state $\brho_i$. This matrix measures the deviations of the stability matrix diagonal elements $\bar{\stability_i} = \stability - \Tr(\stability\brho_i)\boldsymbol{\mathbb{1}}$ from the $i^\text{th}$ instantaneous Lyapunov exponent $\Tr(\stability\brho_i)$~\cite{DasGreen2022}.
%\sdas{This operator $\logder_{p}$ leads to a classical analogue of the quantum Fisher information that is computable for differentiable dynamical systems~\cite{Sahbani2023classical,DasGreen2023speed}. }

%\sdc{We should check if the inclusion of this prefactor 2 changes any of the following or earlier expression(s).}
%\jrgc{Based on our Slack discussion, let's not add the 2.}

%For this state, the deviation from the stability matrix $\bar{\stability_i} = \stability - \Tr(\stability\brho_i)$ includes the $i^\text{th}$ instantaneous Lyapunov exponent $\Tr(\stability\brho_i)$~\cite{DasGreen2022}. 

To derive Jacobi's formula for a maximally mixed state, consider a complete set of $d$ basis vectors that span the state space. Each of these vectors corresponds to a pure perturbation $\brho_i$ and a propagator
$\widetilde\jacobian_i$ that evolves according to Eq.~\ref{Eq:det-pure}. We sum over all of these equations to get
%\msc{This is a very special case when the number $k$ of the pure state $\brho_i$ equals $d$ the dimension of the identity $\mathbb{1}_d$, or the dimension of the eigenspace. The general case is in Eq.~\ref{eq:jcc9}, but summing over $k$.}
\begin{equation}
\begin{aligned}
   \sum_{i=1}^{d}\frac{d}{dt}\ln|\widetilde{\jacobian}_i| &= \sum_{i=1}^{d}(\Tr\stability - d \Tr(\stability\brho_i)) \\
         &= d\left(\Tr\stability - \sum_{i=1}^d \Tr(\stability\brho_i)\right).
\end{aligned}
\end{equation}
Here $|\widetilde{\jacobian}'|$ is the determinant of the product: $|\widetilde{\jacobian}'|=\left(\prod_{i=1}^{d}|\widetilde{\jacobian}_i|\right)^{1/d}$. We use this expression to simplify the left side by substituting $\sum_{i=1}^d \ln |\widetilde{\jacobian}_i'|= \frac{1}{d} \ln |\widetilde\jacobian'|$. The determinant $|\widetilde{\jacobian}'|$ then scales the phase space volume by the action of the $d$ propagators $\widetilde{\jacobian}'_i$:
\begin{align}\label{Eq:det-max-mix}
\Lambda_{\text{max}} := \frac{d}{dt}\ln |\widetilde{\jacobian}'| &= \Tr\stability - d\Tr(\stability\brho_{\text{max}})=0, 
\end{align}
To get $\brho_\text{max}$ on the right, we used $\sum_{i=1}^d\brho_i = d\sum_{i=1}^d\frac{1}{d}\brho_i = d\brho_{\text{max}}$. 
This Jacobi formula for the maximally mixed state defines another contraction rate, $\Lambda_{\text{max}}$. 
If there was one basis state, this rate would reduce to $\Lambda_p$. 
However, $\Lambda_{\text{max}}$ vanishes because the set of basis states is complete.

%where we define a \textit{maximally mixed} state by summing over all the basis states $\brho_{\text{max}} = d^{-1}\sum_{i=1}^d \brho_i$. 
%an extension of Eq.~\ref{Eq:det-pure}.

We can interpret this zero contraction rate as another form of Liouville's equation~\cite{DasGreen2022}.
First, we notice that the right-hand side of Eq.~\ref{Eq:det-max-mix} measures the deviation of phase space compressibility $\Lambda$ of the linear system in  Eq.~\ref{eq:linear-transformation} by the factor $d\Tr(\stability\brho_{\text{max}})$.
This amount of deviation is proportional to the expectation value of the stability matrix $\stability$ with respect to the maximally mixed state $\brho_{\text{max}}$, and it can vanish in some special cases.
The deviation is accounted for by the term $\Tr(\stability\brho)$, which is related to the phase space compressibility rate per state space dimension. We can see this by expressing the phase space compressibility $\Lambda$ as $\Tr(\stability\brho_{\text{max}}) = \sum^d_{i=1}\Tr(\stability\brho_i)/d = \Lambda/d$ if $\sum^d_{i=1}\brho_i = \boldsymbol{\mathbb{1}}$ for a maximally mixed state $\brho_{\text{max}}$. 
Here the contraction rate is $\sum^d_{i=1}\Tr(\stability\brho_i)$, the sum of all instantaneous Lyapunov exponents~\cite{benettin1980lyapunov, kargin2008lyapunov}. 
Thus, the right-hand-side of Eq.~\ref{Eq:det-max-mix} vanishes due to the relation $\Lambda = \Tr\stability$ and we see that $|\widetilde\jacobian'|$ is time-invariant:
%\begin{align}
%\frac{d}{dt}\ln|\widetilde{\jacobian}'| = 0.
%\end{align}
This invariance holds for arbitrary deterministic dynamics. % and is a form of the Liouville's equation for both Hamiltonian and non-Hamiltonian systems.

The determinant of $\brho_{\text{max}}$ also defines a phase space volume that is preserved over time:
\begin{align}\label{eq:max-min}
  \Tr\left(\frac{d}{dt}\ln\brho_{\text{max}}\right) = \frac{d}{dt} \ln |\brho_{\text{max}}|= 0.
\end{align}
In other words, the linear transformation $\brho_{\text{max}}$ always scales the volume of the phase space by the same factor.
In simpler terms, the transformation consistently expands or contracts the area/volume by a fixed amount.
This interpretation is analogous to Liouville's equation, which states that in a Hamiltonian system, the density of points in phase space remains constant over time, meaning that as a system evolves, the ``volume'' occupied by a collection of points in phase space is conserved.
%This interpretation makes Eq.~\ref{eq:max-min} an analogous statement of Liouville's equation in terms of a classical density matrix for non-Hamiltonian systems.

Another way to see these results for the maximally state is when the basis vectors are orthonormal.
From the completeness relation $\sum_{l=1}^{d}\dyad{\delta \boldsymbol{u}_l}=\boldsymbol{\mathbb{1}}$, we see that $\brho_{\text{max}}(t) = d^{-1}\boldsymbol{\mathbb{1}}$ %\msc{The sentence I crossed after Eq.~\ref{eq:9} is mentioned here.} 
and that Eqs.~\ref{Eq:det-max-mix} and~\ref{eq:max-min} are satisfied:
\begin{align}\label{eq:max-mixed-jacob}
|\widetilde \jacobian'| = \text{constant}, \quad \text{and} \quad |\brho_{\text{max}}| = \text{constant}.
\end{align}
So, for a maximally mixed state, both $\widetilde \jacobian'$ and $\brho_{\text{max}}$ reflect the invariance of phase space volume.
This result is a consequence of the preservation of the norm in the dynamics of the density matrix.

\section{Jacobi's formula for an ensemble of pure states}\label{sec:5}

The previous section showed that maximally mixed density matrices conserve volumes. Here we show that this conservation is because the classical probabilities of the mixed state are uncorrelated with phase space volumes.

A (again, classical) mixed state here is a linear sum of an ensemble of $k$ pure perturbation states $\brho_i$ locally at a phase point (App.~\ref{SM:E5}), which has a spectral decomposition
%\sout{Next, we define the probabilities by projecting the set of pure states on a complete set of $d$ orthonormal basis vectors $\bphi_i = \dyad{\delta\boldsymbol{\ms{\bphi_i}}}$ (such as the eigenvectors of $\brho$ or $\stability_+$)}
\begin{align}\label{eq:mx}
\brho(t) = \frac{1}{k} \sum_{i=1}^{k}\brho_i =  \sum_{i=1}^{d}p_i \dyad{\delta \bphi_i}{\delta \bphi_i},
\end{align}
defining the probabilities $p_i = \bra{\delta\bphi_i}\brho\ket{\delta\bphi_i}$. These eigenvectors $\ket{\delta \bphi_i} $ evolve over time by the evolution operator $\widetilde \jacobian_i$. 

To analyze the correlation between volume and the classical probabilities is to first sum Eq.~\ref{eq:jc7} over all the classical probabilities (App.~\ref{SM:E6}), 
\begin{align}\label{eq:jcc9}  
    \sum_{i=1}^{d}p_i{(t)}\frac{d}{dt}{\ln{\lvert  \widetilde \jacobian_i \rvert}^2} = 2\sum_{i=1}^{d}p_i{(t)}\Tr( \bar\stability_i(t)).
\end{align}
Again, the probabilities are for basis states $\ket{\bphi_i}$, which we take to evolve according to the propagator $\widetilde \jacobian_i$ in Eq.~\ref{eq:jc7}. 
We express the left side as
\begin{align}\label{eq:jcc10}
\begin{split}
	\frac{d}{dt}\ln\left(\prod_{i=1}^{d} \lvert  \widetilde \jacobian_i \rvert^{2p_i{(t)}}\right)  &- \sum_{i=1}^{d}\frac{dp_i{(t)}}{dt}\ln{\lvert  \widetilde \jacobian_i \rvert}^2 \\ &=  2  \Tr[\boldsymbol{\stability}(\boldsymbol{\mathbb{1}} - \brho(t){d})]\\
	&=2  [\Lambda - d\Tr(\stability\brho)],
\end{split}
\end{align}
where $\boldsymbol{\mathbb{1}}$ is the $d\times d$ identity matrix. 
%\sdc{It may not be easy for the reader to see how to get the RHS i.e. $2 \Tr[\boldsymbol{\stability}(\boldsymbol{\mathbb{1}}  - \brho(t){d})] $ when are focusing on the LHS here.  It would be helpful to write a sentence or connect to an equation we have used earlier. \msc{We mentioned App.H above, which contains all derivations.}}
Let us define the surprisal $I_i(t) = -\ln p_i(t)$ and the classical average of $\ln{\lvert  \widetilde \jacobian_i \rvert}^2$ as $\langle \ln{\lvert  \widetilde \jacobian \rvert}^2 \rangle= \sum_{i=1}^{d}p_i{(t)}\ln{\lvert  \widetilde \jacobian_i \rvert}^2$, with $\langle\cdot\rangle$ indicating the average with respect to the probability distribution $\{p_i\}$.
Using these definitions, we get the covariance of the derivative of the surprisal (App.~\ref{SM:E7}):
\begin{align}\label{eq:covariance1}
    \operatorname{cov}\left(\dot{I}, \ln{\lvert  \widetilde \jacobian \rvert}^2\right)   =  - \sum_{i=1}^{d}\frac {dp_i{(t)}}{dt}\ln{\lvert  \widetilde \jacobian_i \rvert}^2.
\end{align}
%where the covariance of any two variables $\boldsymbol{x}$ and $\boldsymbol{y}$ is defined as $\operatorname{cov}\left(\boldsymbol{x},\boldsymbol{y}\right) = \langle{\boldsymbol{x}\boldsymbol{y}}\rangle - \langle{\boldsymbol{x}}\rangle\langle{\boldsymbol{y}}\rangle$.
From Eq.~\ref{eq:jcc10}, we obtain the equation of motion
\begin{equation}\label{eq:covariance2}
	\begin{aligned}
	\frac{d}{dt}\langle \ln{\lvert  \widetilde \jacobian \rvert}^2 \rangle + \operatorname{cov}\left(\dot{I}, \ln{\lvert  \widetilde \jacobian \rvert}^2\right) %&=   2  \Tr[\boldsymbol{\stability}(\boldsymbol{\mathbb{1}}  - \brho{d})]\\ 
	&=2  [\Lambda - d\Tr(\stability\brho)]
\end{aligned}
\end{equation}
for the average over the probabilities $p_i$. The evolution of phase space volume over time (averaged over the probabilities) is related directly to the contraction rate and contains the covariance between phase space volumes and the surprisal rates.

For a maximally mixed state $\brho_\text{max} = d^{-1}\boldsymbol{\mathbb{1}}$, the trace in Eq.~\ref{eq:covariance2} vanishes as we saw in the previous section. However, the probabilities are also constant $p_i=1/d$ so that
\begin{align}    
 \operatorname{cov}\left(\dot{I}, \ln{\lvert  \widetilde \jacobian \rvert}^2\right) = 0,
\end{align}
the (statistical) surprisal rate, $\dot{I}_i(t)$, and mechanical $\ln{\lvert  \widetilde \jacobian_i \rvert}^2$ are uncorrelated. 
%A geometric consequence of this lack of correlation is that the maximally mixed density matrix conserves volumes.

\subsection{Recovering Jacobi's formula for pure and maximally mixed states}

%Jacobi's formula and the Liouville equation for both pure and maximally mixed states are special cases of a more general result for mixed states. 
Two special cases show that Eq.~\ref{eq:jcc9} and Eq.~\ref{eq:jcc10} reduce to forms of Liouville's equation (App.~\ref{SM:E8}).

First, consider the pure state.
In this case, the probabilities are all zeros except one probability that equals one.
From Eq.~\ref{eq:jcc9}, we get Jacobi's formula for the pure state $\brho_i$:
%\begin{align}
%    \frac{d\ln\left( \lvert  \widetilde \jacobian_i \rvert^2\right)}{dt} =      2\Tr( \bar\stability_i(t)),
%\end{align}
%which can be written as:
\begin{align}\label{eq:jp}
    \frac{d}{dt}\ln\lvert  \widetilde \jacobian_i \rvert = \Tr( \bar\stability_i)
\end{align}
and recover Eq.~\ref{eq:jc8}. 

Second, consider the maximally mixed state
\begin{align}
	\brho_\text{max} = d^{-1} \sum_{i=1}^{d}\bphi_i =d^{-1}\boldsymbol{\mathbb{1}},
\end{align}
as a linear sum of $d$ basis states $\bphi_i$. 
%For this state, all basis states have equal probabilities $1/d$. 
Equation~\ref{eq:jcc10} then takes the form
%\sdas{\sout{We can discuss some special cases; If the probabilities $p_i(t) = \frac{1}{N}$, the fully mixed state turns to be a maximally mixed state,  $\brho(t) = \frac{1}{N}\sum_{i=1}^{N}\dyad{\delta{u_i(t)}}{\delta{u_i(t)}}$ formed of $N$ pure states $ \dyad{\delta{u_i(t)}}{\delta{u_i(t)}}$ (Appendix~\ref{SM:E6}). The equation of motion in Eq.~\ref{eq:jcc10} leads to 
\begin{align} \label{eq:jcc13}
    \frac{d}{dt}\ln\left(\prod_{i=1}^{d} \lvert  \widetilde \jacobian_i \rvert\right)     =  d  \Tr[\boldsymbol{\stability}(\boldsymbol{\mathbb{1}}  - \brho{_\text{max}}{d})] = 0.
\end{align}
This rate vanishes because $\Tr \stability = d\Tr(\stability \brho{_\text{max}})$ for the maximally mixed state. 
%In general, $\brho(t) = \frac{1}{N}\sum_{i=1}^{N} \dyad{\delta{u_i(t)}}{\delta{u_i(t)}} \neq \frac{\boldsymbol{\mathbb{1}}}{d}$. When the probabilities are constant, $p_i =\frac{\mathbb{1}}{d}$. We get a very remarkable case of the maximally mixed state that is constructed from the uniform standard basis $\dyad{i}{i}$,  $\brho(t) = \frac{\boldsymbol{\mathbb{1}}}{d}$. As a result, the RHS of Eq.~\ref{eq:jcc13} vanishes and leads to the following equation of motion:
%\begin{align}  
%    \frac{d}{dt}\ln\left(\prod_{i=1}^{d} \lvert  \widetilde \jacobian_i \rvert\right)     = 0.
%\end{align}
Therefore, the product of the determinants of the propagator of each state is time independent
\begin{align} \label{eq:jcc14}
      \prod_{i=1}^{d} \lvert  \widetilde \jacobian_i \rvert   = |\widetilde \jacobian'|=  \text{constant},
\end{align}
and we recover Eq.~\ref{eq:max-mixed-jacob}.
If we choose a complete set of the standard basis that does not evolve over time, the propagator is just the identity matrix, $\widetilde \jacobian_i = \boldsymbol{\mathbb{1}}$.
The product of the determinants is $\prod_{i=1}^{d} \lvert  \widetilde \jacobian_i \rvert =  1$.
This is a direct consequence of Eq.~\ref{eq:jcc14} because the determinant of the identity $\widetilde \jacobian_i = \boldsymbol{\mathbb{1}}$ equals one, $\lvert  \widetilde \jacobian_i \rvert = 1$, thus $ \prod_{i=1}^{d} \lvert  \widetilde \jacobian_i \rvert = 1$.
As a result, the maximally mixed state preserves the norm of any vector and the volume of any region of the tangent space.
 
\section{Jacobi's formula for mixed states}\label{sec:6}

We have seen how the well-known phase space contraction rate is a part of the density matrix theory, which raises a question about what this rate, and Jacobi's formula, is for mixed states. %In this section, we derive Jacobi's formula for a mixed state. 
To derive this contraction rate, we start with a mixed state along a classical trajectory in a $d$-dimensional state space, $\brho = \brho(t) = \sum_{i=1}^d p_i(t) \dyad{\delta \bphi_i(t)}$,
%\begin{align}\label{eq:jccc}
%  \brho(t) =  \sum_{i=1}^d p_i \bphi_i.
%\end{align}
%\sdc{The expression above is confusing.  We can represent the mixture of $k$ pure sates in the $d$ basis of $\brho$. We should perhaps $p$'s only if we are using $\brho$ eigenbasis.}
%\begin{align}
%\brho = \frac{\bxi}{\Tr\bxi} 
%= (\Tr\bxi)^{-1}\sum_{i=1}^N \dyad{\delta\ex_i}{\delta\ex_i}
%= (\Tr\bxi)^{-1}\sum_{i=1}^N \bxi_i.
%\end{align}
%{Don't seem to need unnormalized density matrix here. Might be easier to just define this using delta u \msc{Fixed}}
%The corresponding normalized mixed density matrix: 
%\begin{align}
%\brho =\frac{\bxi}{\Tr\bxi}.
%\end{align} 
with normalization such that the probabilities always sum to one.
% for which the off diagonal elements are zeros and the diagonal elements are the probabilities
Because $\brho$ is symmetric positive definite, it has a spectral decomposition
\begin{align}\label{eq:jc11}
    \brho = {\boldsymbol{Q}}\bm\Xi{\boldsymbol{Q}}^{-1}
\end{align}
into an orthogonal matrix of the eigenvectors $\boldsymbol{Q}$ of the mixed state $\brho$ and a diagonal matrix $\bm\Xi$.
The diagonal elements of $\bm\Xi$ are the non-negative eigenvalues of $\brho$, $\{p_m\}$. 
We expand $\bm\Xi$ in the time-invariant standard basis $\{\ket{\boldsymbol{e}_m}\}$:
\begin{align}
\bm\Xi = \sum_{m=1}^dp_m\dyad{ \boldsymbol{e}_m}.
\end{align}
As a result, the time derivatives of $\brho$ and $\bm\Xi$ are equal (App.~\ref{SM:E9}):
\begin{align}\label{eq:jc12}
    \Tr(\frac{d\brho}{dt}\brho^{-1}) = \Tr(\frac{d\bm\Xi}{dt}\bm\Xi^{-1}).
\end{align}
%Since the mixed state $\brho$ is positive definite, its eigenvalues $\{\lambda_k{(t)}\}$ are all positive and it has an inverse, $\brho^{-1}$. The first derivative of the mixed state $\brho$ and the diagonal matrix $\boldsymbol{\Sigma}$, $\boldsymbol{\Sigma} = \sum_{k=1}^{d}\lambda_k{(t)}\ket{k}\bra{k}$ is represented in the standard eigenbasis, with respect to the time parameter leads to the following equality (Appendix~\ref{SM:E6}) 
%\jrgc{Let's find another symbol that's not $\Lambda$. In our other papers, that's a different quantity.}\msc{What about $\Sigma$?}
%\begin{align}\label{eq:jc12}
   % \Tr(\frac{d\brho(t)}{dt}\brho^{-1}) = \Tr(\frac{d\boldsymbol{\Sigma}}{dt}\boldsymbol{\Sigma}^{-1} )
%\end{align}
This connection between $\brho(t)$ and $\bm \Xi(t)$ leads to Jacobi's formula for mixed perturbation states. We first derive Jacobi's formula for the diagonal matrix $\bm\Xi(t)$ in App.~\ref{SM:E10}:
\begin{align}\label{eq:jc13}
   \frac{d\lvert\boldsymbol{ \bm\Xi}\rvert}{dt} = \lvert\boldsymbol{\bm\Xi}\rvert\Tr(\boldsymbol{\bm\Xi}^{-1}\frac{d\boldsymbol{\bm\Xi}}{dt}).
\end{align}
Since the determinant of the mixed state equals the determinant of the diagonal matrix
%\begin{align}\label{eq:jc14}
$\lvert\brho\rvert = \lvert\boldsymbol{\bm{\Xi}}\rvert$,
%\end{align}
we combine Eq.~\ref{eq:jc12} and Eq.~\ref{eq:jc13} to obtain a general form of Jacobi's formula (Apps.~\ref{SM:E11} and~\ref{SM:E12})
\begin{align}\label{eq:jc15}
	\Lambda_m := \frac{1}{2}\frac{d}{dt}\ln\lvert\brho\rvert = \frac{1}{2}\Tr(\brho^{-1}\frac{d\brho}{dt}) = \frac{1}{2}\Tr\logder,
\end{align}
which defines the contraction rate for mixed states $\Lambda_m$. 
Here, $\Tr\logder = \sum_{n=1}^dd_t\ln p_n$ is the trace of the logarithmic derivative of $\brho$~\cite{parisQuantumEstimationQuantum2009}. The logarithmic derivative $\logder$ for a mixed is a generalization of the logarithmic derivative for pure states $\logder_{p}$ in Eq.~\ref{Eq:det-pure}.

Jacobi's formula for mixed states defines a new rate $\Lambda_m$ for the expansion/contraction of ensembles of phase points. The compressible phase space generally implies external forces or dissipative effects are present that cause the phase space volume to shrink or expand. 
This interpretation extends to Jacobi's formula for mixed states. The contraction rate, however, also has statistical contributions from the probabilities of each eigenvector of the density matrix. 
Equation~\ref{eq:jc15} holds for maximally mixed states, which are time invariant $\Lambda_{\text{max}}=0$, and, for pure states, the determinant $|\brho|$ vanishes because one of its eigenvalues is always zero.
With this interpretation, we can consider possible connections between this statistical-mechanical rate and entropy rates.

\section{Contraction rates and entropy rates}\label{sec:7}

The mixed state contraction rate has an implicit average over an ensemble. We show next that it defines a local Gibbs entropy rate.

In the dynamical systems theory approach to nonequilibrium statistical mechanics~\cite{dorfmanIntroductionChaosNonequilibrium1999,gaspardChaosScatteringStatistical1998}, the phase space contraction rate is related to physical quantities, including the entropy flow rate and transport coefficients.
For example, as a phase space density $\rho(\ex,t)$ evolves according to Liouville's equation, the Gibbs entropy rate is the ensemble average of the phase space contraction rate $\Lambda$~\cite{Andrey1985a,Ruelle1996,Ruelle1997}:
\begin{equation}\label{eq:gibbs-entropy-rate}
\begin{aligned}
	\dot{S}_G/k_B &= \int d\bx \dot{\rho}(\bx,t)\ln\rho(\bx,t)\\
				  &= \int d\bx \rho(\bx,t)\Lambda(\bx,t) = +\langle \Lambda\rangle_{\rho}\\
				  &= n\langle\operatorname{Tr} \left(\boldsymbol{A}_{+} \boldsymbol{\varrho}_{\text{max}}\right)\rangle_{\rho}
\end{aligned}
\end{equation}
with averages $\langle\cdot\rangle_{\rho}$ over $\rho$ and $\stability_+ = (\stability + \stability^\top)/2$ being the symmetric part of the stability matrix $\stability$. 
The last line expresses the contraction rate within the density matrix theory~\cite{DasGreen2022,DasGreen2023speed,DasGreen2024} using the maximally mixed states $\brho_{\text{max}}$. 
This connection between one density matrix and an entropy rate suggests that fully mixed states may also have some relation to the rate of expansion and compression of phase space.
However, we would expect this contraction rate to instead be the one that follows from Jacobi's equation for mixed states in Eq.~\ref{eq:jc15}, not $\Lambda$. 

The Gibbs entropy rate depends on the global probability distribution, which can spread over the available phase space. 
While the maximally mixed state relates to $\dot{S}_G$ in Eq.~\ref{eq:gibbs-entropy-rate}, we still need the dynamics of the entire distribution for calculations. However, the mixed state density matrix is a representation of the distribution in local phase space volumes, so we expect the contraction rates here to also define a local entropy rate. 

One possible entropy is the classical von Neumann entropy $\mathcal{S}_V = -\Tr\left(\brho\ln\brho\right) = -\sum_{n=1}^k p_n \ln p_n$.
This entropy has a rate $\dot{\mathcal{S}}_V = -\sum_{n=1}^k d_tp_n \ln p_n$ that depends only on the classical probabilities.
However, another form of the Gibbs entropy rate is directly related to the expansion/contraction of a fully mixed state. 

A specific form of the Gibbs entropy rate follows by identifying the mixed state as a covariance matrix for $k$ variables. 
We can motivate this expression by recalling that for a vector of random variables $\boldsymbol{y}$, the normal distribution is
\begin{equation}
	\rho_G(\by;\by_0) = \frac{\text{exp}(-\frac{1}{2}(\by - \by_0)^\top \bm\chi^{-1}(\by - \by_0))}{\sqrt{(2\pi)^k|\bm\chi|}}
\end{equation}
with mean $\by_0$ and covariance matrix $\boldsymbol{\chi}=\langle\left(\by-\by_0\right)\left(\by-\by_0\right)^{\top}\rangle$.
Now, consider the distribution to be localized in a small volume of phase space and take $\by-\by_0$ to be $\ket{\delta\ex}$. We can similarly translate the sample covariance $\boldsymbol{\chi}=k^{-1}\sum_i^k\left(\by_i-\by_0\right)\left(\by_i-\by_0\right)^{\top}$ to the mixed state density matrix $\brho = k^{-1}\sum_i^k\dyad{\delta\ex}$.
In other words, the mixed state $\brho$ is akin to a covariance matrix that measures the spread to $k$ phase points $\ex_i$ projected on a unit hypersphere centered at a given phase point $\ex_0$ at time $t_0$. 
Since $|c\brho| = c^d |\brho|$, the local Gibbs (differential Shannon) entropy is (App.~\ref{SM:E13})
\begin{equation}
\begin{aligned}
	\mathcal{S}_G %=-k_B\int\mathcal{W}\ln\mathcal{W}\,{d\ex}
	&=  -k_B\int \rho_G(\ex)\ln\rho_G(\ex)\,{d\ex} \\
	&=\frac{k_B}{2} \ln\left(\left(2\pi e c\right)^d|\brho|\right)
\end{aligned}
\end{equation}
using the typeface $\mathcal{S}_G$ to indicate this local form of $S_G$. 
This entropy, $\mathcal{S}_G$, quantifies the dispersion and correlation among the statistical mixture of normalized perturbation vectors in a small volume of phase space. 
The time derivative is 
\begin{equation}\label{eq:gibbs2}
	\dot{\mathcal{S}}_G/k_B %= \frac{1}{2}\frac{d}{dt} \ln |\bm\chi| 
	= \frac{1}{2}\frac{d}{dt} \ln |\brho|
	=  \frac{1}{2}\sum_{i=1}^{d} \frac{d}{dt} \ln p_i,
\end{equation}
where the last expression represents $\brho$ in its eigenbasis. %, $\brho = \sum_{i = 1}^{d} p_i \boldsymbol{\bphi}_i$ so that $|\brho| = \prod_{i=1}^{d} p_i$ is in terms of the eigenvalues $p_i$.
Stepping back, we recognize this entropy rate is set by Eq.~\ref{eq:jc15} -- the analog of Jacobi's formula for mixed states. 
%The Gibbs entropy rate $\dot{\mathcal{S}}_G$ in Eq.~\ref{eq:gibbs2} is
So, we can rewrite the Gibbs entropy rate here
\begin{equation}
	\dot{\mathcal{S}}_G/k_B = \frac{1}{2}\Tr \logder = \Lambda_m
\end{equation}
to show that it is still directly proportional to a ``contraction rate''. However, in this case the contraction rate is the one derived from Jacobi's formula for mixed states, $\Lambda_m$. %, Eq.~\ref{eq:jc15}.

Andrey was the first to recognize that the Gibbs entropy rate is the average phase space contraction rate $\dot{S}_G/k_B=\langle{\Lambda}\rangle$~\cite{Andrey1985a}. 
Here we find when using classical mixed states, which represent the distribution in local phase space volumes, that the local Gibbs entropy rate is proportional to another phase space contraction rate $\dot{\mathcal{S}}_G/k_B=\Lambda_m$.
Neither $\Lambda$ nor $\Lambda_m$ has a definite sign, but if there is a \textit{net} contraction of the phase space associated with $\brho$, then these rates are negative and the system is dissipative.
In this case, we would associate $\langle \Lambda\rangle_\rho$~\cite{Klages2007} and $\Lambda_m$ with the entropy flow rate, the entropy dissipated by the system to the surroundings. 
The entropy flow rate $-\dot{\mathcal{S}}_G$ here avoids Liouville's equation for the probability density here, $d_t\ln\rho(\ex,t)= -\Lambda$.
Instead, we have ``encoded'' the statistical evolution of probability directly in the dynamics of the mixed state density matrix.
This theoretical choice gives the density matrix an advantage that is in part conceptual, in that one can draw directly from quantum dynamics, and in part computational, in that the density matrix representing the ensemble of phase points is readily computable by numerically solving its equation of motion. %~\cite{DasSG2025}.

\section{Conclusions}

To summarize, we have extended a classical density matrix for the statistical mechanics of non-Hamiltonian systems. 
The theory is based on a modification of tangent space dynamics to preserve the norm of tangent vectors and redefine the state of dynamical systems to include pure (from single tangent vectors), maximally mixed (from a complete basis of tangent vectors), or mixed density matrices (from a statistical mixture of tangent vectors). 
In each of these cases, we derived Jacobi's formula for the propagator and density matrix, giving a consistent set of definitions for the corresponding contraction rates. 
These rates all have forms that mirror the usual Jacobi formula and can also be expressed in terms of their respective logarithmic derivatives.
For maximally mixed states, the phase space volume contraction rate $\Lambda_{\text{max}}$ vanishes and Jacobi's formula is analogous to Liouville's equation for conservative systems. 
Jacobi's formula for pure and mixed states defines contraction rates $\Lambda_p$ and $\Lambda_m$ that are analogous to those traditionally used to measure phase space compressibility in non-Hamiltonian dynamics. 
These new rates allow us to analyze the phase space compressibility of classical dynamical systems using the density matrix theory. 
With this approach, we do not need to separately account for the dynamics of the probability density; these dynamics are implicit in the evolution of the mixed state density matrix, which we can interpret as a covariance matrix. 
Moreover, because a mixed state represents an ensemble of trajectories, its contraction rate is directly related to the Gibbs entropy rate and the exchange of entropy between a system and its surroundings. 

\begin{acknowledgments}

This material is based upon work supported by the National Science Foundation under Grant No.~2124510.

\end{acknowledgments}

%\bibliography{references}


\clearpage
%\setcounter{figure}{0}
%\renewcommand{\thefigure}{SM\arabic{figure}}
%\title{Supplemental Material: Classical density matrix and Jacobi's formula for non-Hamiltonian systems}
%\maketitle
\onecolumngrid

\appendix

\setcounter{section}{0}

\section{Equation of motion for the propagator}\label{SM:E1}

A tangent vector $\ket{\delta{\yu}(t)}$ evolves according to
\begin{align}\label{eq:jc3}
    \ket{\delta{\yu}(t)} = \widetilde \jacobian(t,t_0)\ket{\delta{\yu}(t_0)}.
\end{align}
The time derivative of this equation gives 
\begin{align}
 \ket{\delta{\dot{\yu}}(t)} = \frac{d \widetilde \jacobian(t,t_0)}{dt}\ket{\delta{\yu}(t_0)} = \frac{d \widetilde \jacobian(t,t_0)}{dt} \widetilde \jacobian^{-1}(t,t_0)\ket{\delta{\yu}(t)}.
\end{align}
Comparing to Eq.~\ref{eq:EOM-unit}, we can identify
\begin{align} 
 \frac{d \widetilde \jacobian(t,t_0)}{dt} \widetilde \jacobian^{-1}(t,t_0) =  \bar\stability(t),
\end{align}
which is Eq.~\ref{eq:jc4}:
\begin{align} 
 \frac{d \widetilde \jacobian(t,t_0)}{dt}  =  \bar\stability(t){\widetilde \jacobian(t,t_0)}.
\end{align}

\section{Propagator properties}\label{SM:E2}

\textit{Time locality.--} The tangent vector $\ket{\delta{\yu}(t)}$ evolves according to
\begin{align}
    \ket{\delta{\yu}(t)} =  \widetilde \jacobian(t,t_0)\ket{\delta{\yu}(t_0)}.
\end{align}
Several properties of the propagator $\widetilde\jacobian$ follow from the choice of the initial and final times. 
At $t = t_0$, we see that
\begin{align} 
    \ket{\delta{\yu}(t_0)} =  \widetilde \jacobian(t_0,t_0)\ket{\delta{\yu}(t_0)},
\end{align}
and we can conclude that $\widetilde \jacobian(t_0,t_0) = \boldsymbol{\mathbb{1}}$.

\textit{Composition.--}
In two successive time intervals $[t_0,t_1]$ and $[t_1,t_2]$, we see that
\begin{equation} 
\begin{aligned} 
  \ket{\delta{\yu}(t_2)} &= \widetilde \jacobian(t_2,t_1)\ket{\delta{\yu}(t_1)}\\
  \ket{\delta{\yu}(t_1)} &= \widetilde \jacobian(t_1,t_0)\ket{\delta{\yu}(t_0)}
\end{aligned}
\end{equation}
and combining these
\begin{align} 
  \ket{\delta{\yu}(t_2)} =  \widetilde \jacobian(t_2,t_1)  \widetilde \jacobian(t_1,t_0)\ket{\delta{\yu}(t_0)}.
\end{align}
However, for the entire interval $[t_0,t_2]$, the tangent vector evolves as
\begin{align} 
  \ket{\delta{\yu}(t_2)} =  \widetilde \jacobian(t_2,t_0)\ket{\delta{\yu}(t_0)}
\end{align}
so the propagator has the composition property:
\begin{align} 
     \widetilde \jacobian(t_2,t_0) =  \widetilde \jacobian(t_2,t_1)  \widetilde \jacobian(t_1,t_0).
\end{align}

\textit{Time-reversal.--}
If we take $t_2 = t_0$ in the composition property above, we get
\begin{align} 
    \widetilde \jacobian(t_0,t_0) =  \widetilde \jacobian(t_0,t_1)  \widetilde \jacobian(t_1,t_0) = \boldsymbol{\mathbb{1}}.
\end{align}
Multiplication by $\widetilde\jacobian(t_1,t_0)$ from the left and $\widetilde\jacobian(t_0,t_1)$ from the right gives
\begin{align}
     \widetilde \jacobian(t_1,t_0)  \widetilde \jacobian(t_0,t_1)  \widetilde \jacobian(t_1,t_0)  \widetilde \jacobian(t_0,t_1) =   \widetilde \jacobian(t_1,t_0)   \widetilde \jacobian(t_0,t_1).
\end{align}
We then see the propagator has the time-reversal property:
\begin{align}
    \widetilde \jacobian(t_1,t_0)  \widetilde \jacobian(t_0,t_1) = \boldsymbol{\mathbb{1}}.
\end{align}
We can also matrix multiply the inverse from the left
\begin{align}
    \widetilde \jacobian^{-1}(t,t_0) \widetilde \jacobian(t_1,t_0)  \widetilde \jacobian(t_0,t_1) =  \widetilde \jacobian^{-1}(t,t_0)  
\end{align}
and, because $\widetilde \jacobian^{-1}(t,t_0) \widetilde \jacobian(t_1,t_0) = \boldsymbol{\mathbb{1}}$, we get
\begin{align}
     \widetilde \jacobian^{-1}(t,t_0) =  \widetilde \jacobian(t_0,t).
\end{align}
These properties are given in the main text.

\section{Logarithmic derivative}\label{SM:Logarithmic-derivative}

In this appendix, we derive an expression for the logarithmic derivative $\boldsymbol{L}$. We start from the Lyapunov equation for the mixed state $\brho$:
\begin{align}
    d_t\brho = \frac{ \boldsymbol{L}\brho  +  \brho  \boldsymbol{L}^\top}{2}.
\end{align}
Using the eigendecomposition $\brho = \sum_{i=1}^{d}p_{i}\left( t\right)\dyad{\delta \bphi_i(t)}$, where $\sum_{i=1}^{d}\dyad{\delta \bphi_i} = \boldsymbol{\mathbb{1}}$ and $\brho \ket{\delta \bphi_i}= p_{i}\ket{ \delta \bphi_i}$, we get
\begin{align}
\begin{split}
    2d_t\brho  &= \sum_{i=1}^{d} \dyad{ \delta \bphi_i} \boldsymbol{L}\sum_{j=1}^{d} \dyad{ \delta \bphi_j}\brho  +  \brho\sum_{i=1}^{d} \dyad{ \delta \bphi_i} \boldsymbol{L}^\top\sum_{j=1}^{d}\dyad{ \delta \bphi_j}\\
    &= \sum_{i,j=1}^{d} \dyad{ \delta \bphi_i} \boldsymbol{L}\dyad{ \delta \bphi_j}\brho  +   \sum_{i,j=1}^{d}\brho  \dyad{ \delta \bphi_i} \boldsymbol{L}^\top\dyad{ \delta \bphi_j}\\
    &= \sum_{i,j=1}^{d} \dyad{ \delta \bphi_i} \boldsymbol{L}\dyad{ \delta \bphi_j}p_{j} +   \sum_{i,j=1}^{d}p_{i} \dyad{ \delta \bphi_i} \boldsymbol{L}^\top\dyad{ \delta \bphi_j}\\
    &= \sum_{i,j=1}^{d}\left(p_{j} \bra{ \delta \bphi_i} \boldsymbol{L}\ket{ \delta \bphi_j} + 
 p_{i}\bra{ \delta \bphi_i} \boldsymbol{L}^\top\ket{ \delta \bphi_j}\right) \dyad{ \delta \bphi_i}{ \delta \bphi_j}
\end{split}
\end{align}
We conclude that
\begin{align}
    2\bra{ \delta \bphi_i}d_ t\brho \ket{ \delta \bphi_j} =  p_{j} \bra{ \delta \bphi_i} \boldsymbol{L}\ket{ \delta \bphi_j} + 
 p_{i}\bra{ \delta \bphi_i} \boldsymbol{L}^\top\ket{ \delta \bphi_j}.
\end{align}
The logarithmic derivative can be split to a symmetric part $\boldsymbol{L}_{+}$ and an anti-symmetric part $\boldsymbol{L}_{-}$. Plugging in $\boldsymbol{L}=\boldsymbol{L}_{+}+\boldsymbol{L}_{-}$ in the above expression, and using $\boldsymbol{L}_{+}^\top=\boldsymbol{L}_{+}$ and $\boldsymbol{L}_{-}^\top=-\boldsymbol{L}_{-}$, leads to
\begin{align}
\begin{split}  2\bra{ \delta \bphi_i}d_t\brho\ket{ \delta \bphi_j} &=  p_{j} \bra{ \delta \bphi_i}  \left(\boldsymbol{L}_{+}+\boldsymbol{L}_{-}\right)\ket{ \delta \bphi_j} + 
 p_{i}\bra{ \delta \bphi_i}  \left(\boldsymbol{L}_{+}+\boldsymbol{L}_{-}\right)^\top\ket{ \delta \bphi_j}\\
 &= \left(p_{j} + p_{i}\right)\bra{ \delta \bphi_i} \boldsymbol{L}_{+}\ket{ \delta \bphi_j} + \left(p_{j} - p_{i}\right)\bra{ \delta \bphi_i} \boldsymbol{L}_{-}\ket{ \delta \bphi_j}
\end{split}
\end{align}
Rearranging gives
\begin{align}
   \bra{ \delta \bphi_i} \boldsymbol{L}_{+}\ket{ \delta \bphi_j} = 2\frac{\bra{ \delta \bphi_i}d_t\brho\ket{ \delta \bphi_j}}{p_{i} + p_{j}} +\frac{ p_{i} - p_{j} }{p_{i} + p_{j}}\bra{ \delta \bphi_i} \boldsymbol{L}_{-}\ket{ \delta \bphi_j}. 
\end{align}
Replacing the entries of the symmetric part, $\boldsymbol{L}_{+}=\boldsymbol{L}-\boldsymbol{L}_{-}$, yields  
\begin{align}
\begin{split}
   \bra{ \delta \bphi_i} \boldsymbol{L}\ket{ \delta \bphi_j} &= 2\frac{\bra{ \delta \bphi_i}d_t\brho\ket{ \delta \bphi_j}}{p_{i} + p_{j}} +\left(\frac{ p_{i} - p_{j} }{p_{i} + p_{j}}+1\right)\bra{ \delta \bphi_i} \boldsymbol{L}_{-}\ket{ \delta \bphi_j}\\
   &= 2\frac{\bra{ \delta \bphi_i}d_t\brho\ket{ \delta \bphi_j}}{p_{i} + p_{j}} + \frac{ 2p_{i} }{p_{i} + p_{j}} \bra{ \delta \bphi_i} \boldsymbol{L}_{-}\ket{ \delta \bphi_j}.
\end{split}
\end{align}
Thus, we arrive at the expression of the logarithmic derivative
\begin{align}
      \boldsymbol{L}  = 2\sum_{i,j=1}^{d}\frac{\bra{ \delta \bphi_i}d_t\brho\ket{ \delta \bphi_j}}{p_{i} + p_{j}}\dyad{ \delta \bphi_i}{ \delta \bphi_j} + \sum_{i,j=1}^{d}\frac{ 2p_{i} }{p_{i} + p_{j}} \bra{ \delta \bphi_i} \boldsymbol{L}_{-}\ket{ \delta \bphi_j}\dyad{ \delta \bphi_i}{ \delta \bphi_j},
\end{align}
which is Eq.~\ref{eq: Logarithmic-derivative}.

\section{Symmetric logarithmic derivative}\label{SM: Symmetric-Logarithmic-derivative}
We expand the expression for elements of $d_t\brho$ to find:
\begin{align}
\begin{split}
\bra{ \delta \bphi_i}d_t\brho\ket{ \delta \bphi_j} &= \bra{ \delta \bphi_i}\left(\sum_{k=1}^{d}\frac{dp_{k}}{dt} \dyad{ \delta \bphi_k} +  \sum_{k=1}^{d}p_{k} \dyad{d_t{ \delta \bphi_k}}{ \delta \bphi_k} + \sum_{k=1}^{d}p_{k} \dyad{ \delta \bphi_k}{d_t{ \delta \bphi_k}} \right)\ket{ \delta \bphi_j}\\
&=   \sum_{k=1}^{d}\frac{dp_{k}}{dt}\bra{ \delta \bphi_i}\ket{ \delta \bphi_k}\bra{ \delta \bphi_k}\ket{ \delta \bphi_j} +  \sum_{k=1}^{d}p_{k}\bra{ \delta \bphi_i}\ket{d_t{ \delta \bphi_k}}\bra{ \delta \bphi_k}\ket{ \delta \bphi_j} + \sum_{k=1}^{d}p_{k}\bra{ \delta \bphi_i}\ket{ \delta \bphi_k}\bra{d_t{ \delta \bphi_k}}\ket{ \delta \bphi_j}  \\
&=  \sum_{k=1}^{d}\frac{dp_{k}}{dt}\delta_{ik}\delta_{kj} +  \sum_{k=1}^{d}p_{k}\bra{ \delta \bphi_i}\ket{d_t{ \delta \bphi_k}}\delta_{kj}  + \sum_{k=1}^{d}p_{k}\delta_{ik}\bra{d_t{ \delta \bphi_k}}\ket{ \delta \bphi_j} \\
&=   \frac{dp_{j}}{dt}\delta_{ij}  +   p_{j}\bra{ \delta \bphi_i}\ket{d_t{ \delta \bphi_j}}  +  p_{i}\bra{d_t{ \delta \bphi_i}}\ket{ \delta \bphi_j}\\
&=  \frac{dp_{j}}{dt}\delta_{ij}  +   \left(p_{j}-p_{i}\right)\bra{ \delta \bphi_i}\ket{d_t{ \delta \bphi_j}}.
\end{split}
\end{align}
Plugging in this result into our expression for the symmetric logarithmic derivative $\boldsymbol{L}= \boldsymbol{L}_+$, we get
\begin{align}
\begin{split}
     \boldsymbol{L}_+ &= 2\sum_{i,j=1}^{d}\frac{\bra{ \delta \bphi_i}d_t\brho\ket{ \delta \bphi_j}}{p_{i} + p_{j}} \dyad{ \delta \bphi_i}{ \delta \bphi_j}\\
     &= 2\sum_{i,j=1}^{d}\frac{ d_tp_{j}\delta_{ij}  +   \left(p_{j}-p_{i}\right)\bra{ \delta \bphi_i}\ket{d_t{ \delta \bphi_j}}}{p_{i} + p_{j}} \dyad{ \delta \bphi_i}{ \delta \bphi_j}.
\end{split}
\end{align}
Thus,
\begin{align}
     \boldsymbol{L}_+= \sum_{k=1}^{d} \frac{d_t{p_k}}{p_{k}} \dyad{ \delta \bphi_k} + 2\sum_{i\neq{j}}^{d}\frac{   \left(p_{j}-p_{i}\right)}{p_{j} + p_{i}}\bra{ \delta \bphi_i}\ket{d_t{ \delta \bphi_j}} \dyad{ \delta \bphi_i}{ \delta \bphi_j},
\end{align}
which is Eq.~\ref{eq: Symmetric-Logarithmic-derivative}.

\section{Determinant of the propagator}\label{SM:E3}

While time-ordering is important in the propagator $\widetilde\jacobian$ over a time interval as shown in Eq.~\ref{eq:jc5}. However, this ordering is not important in its determinant $|\widetilde\jacobian|$ as shown in Eq.~\ref{eq:jc6}. 
The determinant in Eq.~\ref{eq:jc6} follows from
\begin{align}
\begin{split}
     \lvert  \widetilde \jacobian \rvert &= |\widetilde \jacobian(t,t_0)| \\
     &= \left\lvert\mathcal{T}e^{\int_{t_0}^{t} \bar\stability(t')dt'}\right\rvert\\
     &= \left\lvert{e^{\int_{t_0}^{t} \bar\stability(t')dt'}}\right\rvert\\
     &= e^{\Tr(\int_{t_0}^{t} \bar\stability(t')dt')}\\
     &=  e^{\int_{t_0}^{t}\Tr\left(\bar\stability(t')\right)dt'},
\end{split}
\end{align}
which uses the linearity of the trace and two well known properties of determinants: $\lvert{e}^{\boldsymbol{B}}\lvert = e^{\Tr\left(\boldsymbol{B}\right)}$ and $\lvert{\boldsymbol{B}\boldsymbol{C}} \rvert =  \lvert{\boldsymbol{B}} \rvert \lvert{\boldsymbol{C}} \rvert$. 

\section{Jacobi's formula for pure states}\label{SM:E4}

Squaring the determinant of the propagator and taking the time derivative of its natural logarithm, we get  
\begin{align} 
    \frac{d\ln{{\lvert  \widetilde \jacobian \rvert}^2}}{dt} = 2\Tr( \bar\stability(t))
\end{align}
%Which is 
%\begin{align*}
%    \frac{\frac{d{\lvert  \widetilde \jacobian \rvert}}{dt}}{\lvert  \widetilde \jacobian \rvert} = \Tr( \bar\stability(t))
%\end{align*}
Rearranging, we get
\begin{align}
    \frac{d{\lvert  \widetilde \jacobian \rvert}}{dt} =\lvert  \widetilde \jacobian \rvert \Tr( \bar\stability(t)).
\end{align}
Using Eq.~\ref{eq:jc4} for $\bar\stability$,
%\begin{align*} 
%   \bar\stability =  \frac{d \widetilde \jacobian(t,t_0)}{dt} \widetilde \jacobian^{-1}(t,t_0) 
%\end{align*}
we find Jacobi's formula
\begin{align}
    %\frac{d{\lvert  \widetilde \jacobian \rvert}}{dt} = \lvert  \widetilde \jacobian \rvert\Tr\left( \widetilde \jacobian^{-1}(t,t_0)\frac{d \widetilde \jacobian(t,t_0)}{dt}\right).
    \frac{d{\lvert  \widetilde \jacobian \rvert}}{dt} = \lvert  \widetilde \jacobian \rvert\Tr\left( \widetilde \jacobian^{-1}\frac{d \widetilde \jacobian}{dt}\right),
\end{align}
which is Eq.~\ref{eq:jc8}.

\section{Mixed states in tangent space}\label{SM:E5}

Let's consider the equation of motion of the tangent vector
\begin{align}\label{eq:jc2}
    \ket{\delta{\dot{\ex}}} = \stability\ket{\delta{\ex}}.
\end{align}
The corresponding normalized tangent vector evolves as
\begin{align}\label{eq:jc1}
    \ket{\delta{\dot{\yu}}} = \bar{\stability}\ket{\delta{\yu}}
\end{align}
with $\ket{\delta{\yu}} = \ket{\delta{\ex}}/\norm{\ket{\delta{\ex}}}$ and $\bar{\stability} = \stability - \bra{\delta{\yu}}\stability\ket{\delta{\yu}}$. 
From these vectors, we construct $\brho^{\prime}(t) = \sum_{i=1}^{k}\dyad{\delta{\yu_i(t)}}  = \sum_{i=1}^{k}\brho_i(t)$ of $k$ pure states $\brho_i = \dyad{\delta{\yu_i}}$. %\jrgc{Update $N$ with out notation from main text. \msc{Fixed}.
As in quantum mechanics, we need to normalize $\brho^{\prime}$ so that
\begin{align}
	\brho(t) = \frac{\brho^{\prime}(t)}{\Tr\left(\brho^{\prime}(t)\right)} = \sum_{i=1}^{k}\frac{1}{\Tr\left(\brho^{\prime}(t)\right)}\dyad{\delta{\yu_i(t)}}.
\end{align}
The pure state is normalized, so $\bra{\delta{\yu_i(t)}}\ket{\delta{\yu_i(t)}} = 1$ and
\begin{equation}
\begin{aligned}
    \Tr\left(\brho^{\prime}(t)\right) &= \Tr\left(\sum_{i=1}^{k} \dyad{\delta{\yu_i(t)}}\right)\\
&=\sum_{i=1}^{k}\Tr\left( \dyad{\delta{\yu_i(t)}}\right)\\ 
&=\sum_{i=1}^{k}\bra{\delta{\yu_i(t)}}\ket{\delta{\yu_i(t)}}\\
%& = \sum_{i=1}^{k} 1\\
& =  k.
\end{aligned}
\end{equation}
The definition of a mixed state is then
\begin{align}
\brho(t) = \frac{1}{k}\sum_{i=1}^{k} \dyad{\delta{\yu_i(t)}}.
\end{align}
We can prove this is a mixed state by calculating the purity:
\begin{equation}
\begin{aligned}
    \Tr\left(\brho^2\right) &=\frac{1}{k^2}\Tr\left(\sum_{i=1}^{k} \dyad{\delta{\yu_i(t)}} \sum_{j=1}^{k}\dyad{\delta{\yu_j(t)}} \right)\\
    &= \frac{1}{k^2}\sum_{ij=1}^{k}\bra{\delta{\yu_i(t)}}\ket{\delta{\yu_j(t)}}\Tr\left( \dyad{\delta{\yu_i(t)}}{\delta{\yu_j(t)}}\right)\\
    &= \frac{1}{k^2}\sum_{i=1}^{k}\bra{\delta{\yu_i(t)}}\ket{\delta{\yu_i(t)}}\Tr\left( \dyad{\delta{\yu_i(t)}}{\delta{\yu_i(t)}}\right) + \frac{1}{k^2}\sum_{i\neq{j}}^{k}\bra{\delta{\yu_i(t)}}\ket{\delta{\yu_j(t)}}\Tr\left( \dyad{\delta{\yu_i(t)}}{\delta{\yu_j(t)}}\right)\\
    &= \frac{1}{k^2}\sum_{i=1}^{k}\left(\bra{\delta{\yu_i(t)}}\ket{\delta{\yu_i(t)}}\right)^2  + \frac{1}{k^2}\sum_{i\neq{j}}^{k}\left(\bra{\delta{\yu_i(t)}}\ket{\delta{\yu_j(t)}}\right)^2 \\
    &=  \frac{1}{k^2}\sum_{i=1}^{k} 1  + \frac{1}{k^2}\sum_{i\neq{j}}^{k}\left( \cos\left(\theta_{ij}\right)\right)^2 \\
     &=  \frac{1}{k^2}k  + \frac{1}{k^2}\sum_{i\neq{j}}^{k}\left( \cos\left(\theta_{ij}\right)\right)^2\\
      &\leq  \frac{1}{k^2}k  + \frac{1}{k^2}\sum_{i\neq{j}}^{k} 1\\
       &=  \frac{1}{k^2}k  + \frac{1}{k^2}\left(k-1\right)^2\\
       &=  \frac{1}{k^2}k  + \frac{1}{k^2}\left(k^2-2k +1\right)\\
       &= 1  + \frac{\left(k - 1\right)}{k^2} 
\end{aligned}
\end{equation}
The inequality follows from the fact that $\bra{\delta{\yu_i(t)}}\ket{\delta{\yu_j(t)}}= \cos\left(\theta_{ij}\right) \leq 1$. 
So, we have
\begin{align}
    \Tr\left(\brho^2\right)   \leq   
  1  + \frac{\left(k - 1\right)}{k^2} 
\end{align}
and $k \geq 1$. Since $\left(k - 1\right)/k^2 \leq 0$, the purity
\begin{align}
    \Tr\left(\brho^2\right) \leq 1
\end{align}
has the upper bound we expect for a mixed state. 
 
\section{Derivation of Jacobi's formula for an ensemble of pure states}\label{SM:E6}

Summing Eq.~\ref{eq:jc7} over all the probabilities, $p_i$, in the $d$ dimensional phase space, we get
\begin{align}
    \sum_{i=1}^{d}p_i\frac{d{\ln{\lvert  \widetilde \jacobian_i \rvert}^2}}{dt} = 2\sum_{i=1}^{d}p_i\Tr( \bar\stability_i),
\end{align}
which can be written as  
\begin{align}%\label{eq:d0}
    \frac{d}{dt}\left(\sum_{i=1}^{d}p_i\ln{\lvert  \widetilde \jacobian_i \rvert}^2\right)  - \sum_{i=1}^{d}\frac {dp_i}{dt}\ln{\lvert  \widetilde \jacobian_i \rvert}^2= 2\sum_{i=1}^{d}p_i\Tr( \bar\stability_i).
\end{align}
Inserting the expression $\bar\stability_i = \stability - \langle\stability\rangle_i = \stability -\Tr\left(\stability\dyad{\delta \bphi_i}\right)\boldsymbol{\mathbb{1}}$ and using the properties of natural logarithms yields
\begin{align}\label{eq:d1}
\begin{split}
    \frac{d}{dt}\ln\left(\prod_{i=1}^{d} \lvert  \widetilde \jacobian_i \rvert^{2p_i}\right)  - \sum_{i=1}^{d}\frac{dp_i}{dt}\ln{\lvert  \widetilde \jacobian_i \rvert}^2  &= 2\sum_{i=1}^{d}p_i\Tr( \bar\stability_i)\\
																																										   &= 2\sum_{i=1}^{d}p_i\Tr[\stability(\boldsymbol{\mathbb{1}}  - \dyad{\delta\bphi_i}d)]\\
    &=  2 \Tr\left(\stability{\sum_{i=1}^{d}p_i}\right) - 2\Tr\left(\stability\sum_{i=1}^{d}p_i\dyad{\delta \bphi_i} d\right).
\end{split}
\end{align}
Since $\brho(t) = \sum_{i=1}^{d}p_i{(t)} \dyad{\delta \bphi_i(t)}$,
\begin{align}\label{eq:d2}
    \frac{d}{dt}\ln\left(\prod_{i=1}^{d} \lvert  \widetilde \jacobian_i \rvert^{2p_i}\right)  - \sum_{i=1}^{d}\frac{dp_i}{dt}\ln{\lvert  \widetilde \jacobian_i \rvert}^2  =  2  \Tr[\boldsymbol{\stability}(\boldsymbol{\mathbb{1}}  - \brho{d})],
\end{align}
%Or in a more compact way
%\begin{align}\label{eq:d3}
    %\frac{d\ln\left(\prod_{i=1}^{d} \lvert  \widetilde \jacobian_i \rvert^{2p_i{(t)}}\right)}{dt}  -   \ln{\left(\prod_{i=1}^{d}\lvert  \widetilde \jacobian_i \rvert^{2\frac{dp_i{(t)}}{dt}}\right)}  =  2  \Tr[\boldsymbol{\stability}(\boldsymbol{\mathbb{1}}  - \brho(t){d})]  
%\end{align}
which is Eq.~\ref{eq:jcc10}. 

\section{Surprisal rate correlation}\label{SM:E7}

Defining the surprisal $I_i(t) = -  \ln\left(p_i(t)\right)$ leads to its classical average: 
\begin{align}
   \langle \dot{I_i}\rangle =-\sum_{i=1}^{d}p_i\frac{d}{dt}\ln\left(p_i\right)  = -\sum_{i=1}^{d}p_i \frac{\frac{dp_i}{dt}}{p_i} =- \frac{d}{dt}\sum_{i=1}^{d}p_i =0, 
\end{align}
since $\sum_{i=1}^{d}p_i =1$. Similarly, the classical average of  $\ln{\lvert  \widetilde \jacobian_i \rvert}^2$ is
\begin{align}
    \langle \ln{\lvert  \widetilde \jacobian\rvert}^2\rangle = \sum_{i=1}^{d}p_i\ln{\lvert  \widetilde \jacobian_i \rvert}^2.
\end{align}
Thus, the covariance of the derivative of the surprisal and $\ln{\lvert  \widetilde \jacobian_i \rvert}^2$ is 
\begin{align}
	\operatorname{cov}\left(\dot{I_i}, \ln{\lvert  \widetilde \jacobian_i \rvert}^2\right) = \langle \dot{I}\ln{\lvert  \widetilde \jacobian_i \rvert}^2\rangle - \langle\dot{I_i}\rangle\langle\ln{\lvert  \widetilde \jacobian_i \rvert}^2\rangle =-\sum_{i=1}^{d}\frac {dp_i}{dt}\ln{\lvert  \widetilde \jacobian_i \rvert}^2
\end{align}
From Eq.\ref{eq:jcc10}, we get the following equation of motion
\begin{align}\label{eq:d4}
    \frac{d}{dt}\langle \ln{\lvert  \widetilde \jacobian \rvert}^2 \rangle +    \operatorname{cov}\left(\dot{I_i}, \ln{\lvert  \widetilde \jacobian_i \rvert}^2\right) =   2  \Tr[\boldsymbol{\stability}(\boldsymbol{\mathbb{1}}  - \brho(t){d})],  
\end{align}
which is Eq.\ref{eq:covariance2}.

\section{Recovering Jacobi’s formula for pure and maximally mixed states}\label{SM:E8}

The maximally mixed state $\brho = d^{-1}\boldsymbol{\mathbb{1}}$ has constant probabilities, $p_i =d^{-1}$. Using Eq.~\ref{eq:d2} yields
\begin{align} 
	\frac{d}{dt}\ln\left(\prod_{i=1}^{d} \lvert  \widetilde \jacobian_i \rvert^{\frac{2}{d}}\right)   =     2  \Tr[\boldsymbol{\stability}(\boldsymbol{\mathbb{1}}  -  \frac{\boldsymbol{\mathbb{1}}}{d}{d})] =0. 
\end{align}
The product of the determinants of the propagator of each state is constant, 
\begin{align} 
      \prod_{i=1}^{d} \lvert  \widetilde \jacobian_i \rvert   =  |\widetilde \jacobian'|=  \text{constant},  
\end{align}
which is Eq.~\ref{eq:jcc14}.

For a pure state $\brho_i$, the probabilities are all zeros except one probability that equals one. Thus, by using Eq.~\ref{eq:jcc10}, there is only one term left
\begin{align}
    \frac{d}{dt}\ln\left( \lvert  \widetilde \jacobian_1 \rvert^{2\times0}\lvert  \widetilde \jacobian_2\ldots \lvert  \widetilde \jacobian_i \rvert^{2\times1}\ldots \rvert^{2\times0}\ldots \lvert  \widetilde \jacobian_d \rvert^{2\times0}\right)  =  2  \Tr[\boldsymbol{\stability}(\boldsymbol{\mathbb{1}}  - \brho_i(t){d})].   
\end{align}
Hence,
\begin{align}
    \frac{d}{dt}\ln\left( \lvert  \widetilde \jacobian_i \rvert^2\right)   =    2\Tr( \bar\stability_i)
\end{align}
simplifies to Jacobi's formula for the pure state
\begin{align}
     \frac{d}{dt}\ln\left( \lvert  \widetilde \jacobian_i \rvert\right)   =  \Tr( \bar\stability_i), 
\end{align}
which is Eq.~\ref{eq:jp}. 
 
\section{Spectral decomposition of the mixed state}\label{SM:E9}

The mixed state $\brho = \brho(t) = \sum_{i=1}^{k}p_i{(t)}\dyad{\delta{\yu_i(t)}}$ is symmetric and positive definite, $\brho > 0$, so it has a spectral decomposition
\begin{align}
    \brho = {\boldsymbol{Q}}\boldsymbol{\bm\Xi}{\boldsymbol{Q}}^{-1}
\end{align}
in terms of an orthogonal matrix $\boldsymbol{Q}$ of eigenvectors and a diagonal matrix $\boldsymbol{\bm\Xi}$ of eigenvalues. 
The time derivative of this representation is 
\begin{align}
   \frac{d\brho}{dt} = \frac{d\boldsymbol{Q}}{dt}\boldsymbol{\bm\Xi}{\boldsymbol{Q}^{-1}} + {\boldsymbol{Q}}\frac{d\boldsymbol{\bm\Xi}}{dt}{\boldsymbol{Q}^{-1}} + {\boldsymbol{Q}}\boldsymbol{\bm\Xi}{\frac{d\boldsymbol{Q}^{-1}}{dt}}.
\end{align}
Multiplying from the right by $\brho^{-1}$,
\begin{align}
     \frac{d\brho}{dt}\brho^{-1} = \left( \frac{d\boldsymbol{Q}}{dt}\boldsymbol{\bm\Xi}{\boldsymbol{Q}^{-1}} + {\boldsymbol{Q}}\frac{d\boldsymbol{\bm\Xi}}{dt}{\boldsymbol{Q}^{-1}} + {\boldsymbol{Q}}\boldsymbol{\bm\Xi}{\frac{d\boldsymbol{Q}^{-1}}{dt}}\right)\left({\boldsymbol{Q}}\boldsymbol{\bm\Xi}{\boldsymbol{Q}}^{-1}\right)^{-1}
\end{align}
we can use the orthogonality of $\boldsymbol{Q}$, $\boldsymbol{QQ}^{-1} = \boldsymbol{\mathbb{1}}$, 
%\begin{align}
%     \frac{d\brho(t)}{dt}\brho^{-1} =  \frac{d\boldsymbol{Q}}{dt}\boldsymbol{\bm\Xi}{\boldsymbol{Q}^{-1}}{\boldsymbol{Q}}\boldsymbol{\bm\Xi}^{-1}{\boldsymbol{Q}}^{-1} + {\boldsymbol{Q}}\frac{d\boldsymbol{\bm\Xi}}{dt}{\boldsymbol{Q}^{-1}}{\boldsymbol{Q}}\boldsymbol{\bm\Xi}^{-1}{\boldsymbol{Q}}^{-1} + {\boldsymbol{Q}}\boldsymbol{\bm\Xi}{\frac{d\boldsymbol{Q}^{-1}}{dt}}{\boldsymbol{Q}}\boldsymbol{\bm\Xi}^{-1}{\boldsymbol{Q}}^{-1}
%\end{align}
%Because $\boldsymbol{Q}$ is orthogonal, $\boldsymbol{QQ}^{-1} = \boldsymbol{\mathbb{1}}$, we get
%\begin{align}
%     \frac{d\brho(t)}{dt}\brho^{-1} =  \frac{d\boldsymbol{Q}}{dt}\boldsymbol{ \bm\Xi\bm\Xi}^{-1}{\boldsymbol{Q}}^{-1} + {\boldsymbol{Q}}\frac{d\boldsymbol{\bm\Xi}}{dt}\boldsymbol{\bm\Xi}^{-1}{\boldsymbol{Q}}^{-1} + {\boldsymbol{Q}}\boldsymbol{\bm\Xi}{\frac{d\boldsymbol{Q}^{-1}}{dt}}{\boldsymbol{Q}}\boldsymbol{\bm\Xi}^{-1}{\boldsymbol{Q}}^{-1}
%\end{align}
and $\boldsymbol{\bm\Xi\bm\Xi}^{-1} = \boldsymbol{\mathbb{1}}$, to get
\begin{align}
     \frac{d\brho}{dt}\brho^{-1} =  \frac{d\boldsymbol{Q}}{dt}{\boldsymbol{Q}}^{-1} + {\boldsymbol{Q}}\frac{d\boldsymbol{\bm\Xi}}{dt}\boldsymbol{\bm\Xi}^{-1}{\boldsymbol{Q}}^{-1} + {\boldsymbol{Q}}\boldsymbol{\bm\Xi}{\frac{d\boldsymbol{Q}^{-1}}{dt}}{\boldsymbol{Q}}\boldsymbol{\bm\Xi}^{-1}{\boldsymbol{Q}}^{-1}
\end{align}
Taking the trace
\begin{align}
    \Tr(\frac{d\brho}{dt}\brho^{-1}) = \Tr(\frac{d\boldsymbol{Q}}{dt}{\boldsymbol{Q}}^{-1}) + \Tr({\boldsymbol{Q}}\frac{d\boldsymbol{\bm\Xi}}{dt}\boldsymbol{\bm\Xi}^{-1}{\boldsymbol{Q}}^{-1}) + \Tr({\boldsymbol{Q}}\boldsymbol{\bm\Xi}{\frac{d\boldsymbol{Q}^{-1}}{dt}}{\boldsymbol{Q}}\boldsymbol{\bm\Xi}^{-1}{\boldsymbol{Q}}^{-1})
\end{align}
%This will be
%\begin{align}
%    \Tr(\frac{d\brho(t)}{dt}\brho^{-1}) = \Tr(\frac{d\boldsymbol{Q}}{dt}{\boldsymbol{Q}}^{-1}) + \Tr({\boldsymbol{Q}}^{-1}{\boldsymbol{Q}}\frac{d\boldsymbol{\bm\Xi}}{dt}\boldsymbol{\bm\Xi}^{-1} ) + \Tr(\boldsymbol{\bm\Xi}^{-1}{\boldsymbol{Q}}^{-1}{\boldsymbol{Q}}\boldsymbol{\bm\Xi}{\frac{d\boldsymbol{Q}^{-1}}{dt}}{\boldsymbol{Q}} )
%\end{align}
%Thus
%\begin{align}
%    \Tr(\frac{d\brho(t)}{dt}\brho^{-1}) = \Tr(\frac{d\boldsymbol{Q}}{dt}{\boldsymbol{Q}}^{-1}) + \Tr(\frac{d\boldsymbol{\bm\Xi}}{dt}\boldsymbol{\bm\Xi}^{-1} ) + \Tr(\frac{d\boldsymbol{Q}^{-1}}{dt}{\boldsymbol{Q}} )
%\end{align}
%Then
%\begin{align}
%    \Tr(\frac{d\brho(t)}{dt}\brho^{-1}) = \Tr(\frac{d\boldsymbol{Q}}{dt}{\boldsymbol{Q}}^{-1} + {\boldsymbol{Q}}\frac{d\boldsymbol{Q}^{-1}}{dt}) + \Tr(\frac{d\boldsymbol{\bm\Xi}}{dt}\boldsymbol{\bm\Xi}^{-1} )
%\end{align}
%Which is 
%\begin{align}
%    \Tr(\frac{d\brho(t)}{dt}\brho^{-1}) = \Tr(\frac{d(\boldsymbol{QQ}^{-1})}{dt}) + \Tr(\frac{d\boldsymbol{\bm\Xi}}{dt}\boldsymbol{\bm\Xi}^{-1} )
%\end{align}
%Thus
and simplifying gives
\begin{align}
    \Tr(\frac{d\brho}{dt}\brho^{-1}) = \Tr(\frac{d\boldsymbol{\mathbb{1}}}{dt}) + \Tr(\frac{d\boldsymbol{\bm\Xi}}{dt}\boldsymbol{\bm\Xi}^{-1} )
\end{align}
Because $d_t\Tr\boldsymbol{\mathbb{1}} =0$, we find
\begin{align}\label{eq:half-jacobi}
    \Tr(\frac{d\brho}{dt}\brho^{-1}) = \Tr(\frac{d\boldsymbol{\bm\Xi}}{dt}\boldsymbol{\bm\Xi}^{-1}),
\end{align}
which is Eq.~\ref{eq:jc12}.

\section{Jacobi's formula for the diagonal state}\label{SM:E10}

Now, represent the diagonal matrix $\boldsymbol{\bm\Xi}$ in the standard basis $\{\ket{\boldsymbol{e}_m}\}$
\begin{align}
 \boldsymbol{\bm\Xi}(t) = \begin{pmatrix}
   p_1(t) & 0 & \dots & 0 \\
    0 & p_2(t) & \dots & 0 \\
    \vdots & \vdots & \ddots & \vdots \\
    0 & 0 & \dots & p_d(t)
  \end{pmatrix}  = \sum_{m=1}^{d}p_m{(t)} \dyad{\boldsymbol{e}_m}
\end{align}
and take its derivative with respect to time (and dropping the explicit time dependence)
 \begin{align}
 \frac{d\boldsymbol{\bm\Xi} }{dt} = \begin{pmatrix}
   \frac{dp_1 }{dt} & 0 & \dots & 0 \\
    0 & \frac{dp_2 }{dt} & \dots & 0 \\
    \vdots & \vdots & \ddots & \vdots \\
    0 & 0 & \dots & \frac{dp_d}{dt}
  \end{pmatrix} = \sum_{m=1}^{d}\frac{dp_m}{dt}\dyad{\boldsymbol{e}_m}.
\end{align}
Multiplication by $\boldsymbol{\Xi}^{-1}$ from the right gives
\begin{align}
 \frac{d\boldsymbol{\bm\Xi}}{dt}\boldsymbol{\bm\Xi}^{-1} =  \left(\sum_{m=1}^{d}\frac{dp_m}{dt} \dyad{\boldsymbol{e}_m}  \right)\left(\sum_{s=1}^{d}p_s^{-1} \dyad{\boldsymbol{e}_s}\right).
\end{align}
We can then conclude that
\begin{align}
 \frac{d\boldsymbol{\bm\Xi}}{dt}\boldsymbol{\bm\Xi}^{-1} = \sum_{m=1}^{d}\frac{dp_m}{dt}p_m^{-1}\dyad{\boldsymbol{e}_m}   
\end{align}
%For simplicity
%\begin{align}
% \frac{d\boldsymbol{\bm\Xi}}{dt}\boldsymbol{\bm\Xi}^{-1} = \sum_{k=1}^{d}\frac{d\lambda_k}{dt}\lambda_k^{-1} \dyad{k}{k}   
%\end{align}
and the trace is just
\begin{align}
 \Tr(\frac{d\boldsymbol{\bm\Xi}}{dt}\boldsymbol{\bm\Xi}^{-1}) =  \sum_{m=1}^{d}\frac{dp_m}{dt}p_m^{-1} = \sum_{m=1}^{d}\frac{d}{dt}\ln(p_m).
\end{align}
We can manipulate the trace into another Jacobi equation
\begin{align}
	\Tr(\frac{d\boldsymbol{\bm\Xi}}{dt}\boldsymbol{\bm\Xi}^{-1}) =  \frac{d}{dt}\sum_{m=1}^{d}\ln(p_m)
 = \frac{d}{dt}\ln(\prod_{m=1}^{d}p_m)
= |\boldsymbol{\bm\Xi}|^{-1}\frac{d|\boldsymbol{\bm\Xi}|}{dt}.
\end{align}
The last equality uses $|\boldsymbol{\bm\Xi}| = \prod_{m=1}^{d}p_m$. 
%This leads to
%\begin{align}
% \Tr(\frac{d\boldsymbol{\bm\Xi}}{dt}\boldsymbol{\bm\Xi}^{-1}) =  \frac{d \ln(|\boldsymbol{\bm\Xi}|)}{dt} = \frac{\frac{d|\boldsymbol{\bm\Xi}|}{dt}}{|\boldsymbol{\bm\Xi}|} 
%\end{align}
%Or,
%\begin{align}
% \Tr(\frac{d\boldsymbol{\bm\Xi}}{dt}\boldsymbol{\bm\Xi}^{-1}) =  \frac{d \ln(\lvert\boldsymbol{\bm\Xi}\rvert)}{dt} = \frac{\frac{d\lvert\boldsymbol{\bm\Xi}\rvert}{dt}}{\lvert\bm\Xi\rvert} 
%\end{align}
Rearranging, we can conclude that
\begin{align}\label{eq:diagonal}
   \frac{d\lvert\boldsymbol{\bm\Xi}\rvert}{dt} = \lvert\boldsymbol{\bm\Xi}\rvert\Tr(\boldsymbol{\bm\Xi}^{-1}\frac{d\boldsymbol{\bm\Xi}}{dt})
\end{align}
is Jacobi's formula for the diagonal matrix $\boldsymbol{\bm\Xi}$.

\section{Jacobi's formula for mixed states}\label{SM:E11}

Jacobi's formula also follows for the density matrix itself. Starting from the spectral decomposition,
\begin{align}
    \brho = {\boldsymbol{Q}}\boldsymbol{\bm\Xi}{\boldsymbol{Q}}^{-1},
\end{align}
we can take the determinant
\begin{align}
    \lvert\brho\rvert = \lvert{\boldsymbol{Q}}\rvert\lvert\boldsymbol{\bm\Xi}\rvert\lvert{\boldsymbol{Q}}^{-1}\rvert = \lvert\boldsymbol{\bm\Xi}\rvert\lvert{\boldsymbol{Q}}\rvert\lvert{\boldsymbol{Q}}^{-1}\rvert = \lvert\boldsymbol{\bm\Xi}\rvert\lvert{\boldsymbol{Q}}{\boldsymbol{Q}}^{-1}\rvert
\end{align}
and find its determinant
\begin{align}\label{eq:density-diagonal}
    \lvert\brho(t)\rvert = \lvert\boldsymbol{\bm\Xi}\rvert\lvert{\boldsymbol{\mathbb{1}}}\rvert = \lvert\boldsymbol{\bm\Xi}\rvert
\end{align}
is identical to the determinant of the diagonal matrix. Taking the derivative of the equation.~\ref{eq:density-diagonal} and combining Eq.~\ref{eq:half-jacobi} and Eq.~\ref{eq:diagonal}, we get
\begin{align}
   \frac{d\lvert\brho\rvert}{dt} = \frac{d\lvert\boldsymbol{\bm\Xi}\rvert}{dt} = \lvert\boldsymbol{\bm\Xi}\rvert\Tr(\boldsymbol{\bm\Xi}^{-1}\frac{d\boldsymbol{\bm\Xi}}{dt}) = \lvert\brho\rvert\Tr(\brho^{-1}\frac{d\brho}{dt})
\end{align}
recognizing Jacobi's formula for the mixed state
\begin{align}\label{eq:jcb1}
   \frac{d\lvert\brho\rvert}{dt} = \lvert\brho\rvert\Tr(\brho^{-1}\frac{d\brho}{dt}),
\end{align}
which is Eq.~\ref{eq:jc15}.
 
\section{Jacobi's formula from the spectral decomposition of the mixed state}\label{SM:E12}

The spectral decomposition of the mixed state is
\begin{align}
 \brho = \boldsymbol{\brho}(t) = \sum_{m=1}^{d}p_m{(t)} \dyad{\boldsymbol{m}(t)} 
\end{align}
where $p_m$ and $\ket{\boldsymbol{m}}$ are the eigenvalues and the eigenvectors of the mixed state $\brho$ in the $d$-dimensional eigenspace. 
Thus,
\begin{align}
 \frac{d\boldsymbol{ \brho}}{dt} =  \sum_{m=1}^{d}\frac{dp_m}{dt} \dyad{\boldsymbol{m}} +\sum_{m=1}^{d}p_n \dyad{d_t{\boldsymbol{m}}}{\boldsymbol{m}}+\sum_{m=1}^{d}p_m \dyad{\boldsymbol{m}}{d_t{\boldsymbol{m}}}
\end{align}
Matrix multiplication gives
\begin{align}
 \frac{d\boldsymbol{ \brho}}{dt}\boldsymbol{ \brho}^{-1} =  \left(\sum_{m=1}^{d}\frac{dp_m}{dt} \dyad{\boldsymbol{m}} +\sum_{m=1}^{d}p_m \dyad{d_t{\boldsymbol{m}}}{\boldsymbol{m}}+\sum_{m=1}^{d}p_m \dyad{\boldsymbol{m}}{d_t{\boldsymbol{m}}}\right)\left(\sum_{l=1}^{d}p_l^{-1} \dyad{\boldsymbol{l}}\right)  
\end{align}
The result
\begin{align}
 \frac{d\boldsymbol{ \brho}}{dt}\boldsymbol{ \brho}^{-1} = \sum_{m=1}^{d}\frac{dp_m}{dt}p_m^{-1} \dyad{\boldsymbol{m}} + \sum_{m=1}^{d} \dyad{d_t{\boldsymbol{m}}}{\boldsymbol{m}} + \sum_{m,l=1}^{d}p_{m}p_l^{-1} \bra{d_t{\boldsymbol{m}}} \ket{\boldsymbol{l}}\dyad{\boldsymbol{m}}{\boldsymbol{l}}
\end{align}
%For simplicity
%\begin{align}
% \frac{d\boldsymbol{ \brho}}{dt}\boldsymbol{ \brho}^{-1} = \sum_{n=1}^{d}\frac{dp_n}{dt}p_n^{-1} \dyad{n}{n} + \sum_{n=1}^{d} \dyad{\partial_t{n}}{n} + \sum_{n,m=1}^{d}p_np_m^{-1} \dyad{n}{\partial_t{n}} \dyad{m}{m}  
%\end{align}
has the trace
\begin{align}
 \Tr(\frac{d\boldsymbol{ \brho}}{dt}\boldsymbol{ \brho}^{-1}) = \sum_{m=1}^{d}\frac{dp_m}{dt}p_m^{-1}  + \sum_{m=1}^{d} \bra{\boldsymbol{m}}\ket{d_t{\boldsymbol{m}}} +  \sum_{m,l=1}^{d} \delta_{lm}\bra{d_t{\boldsymbol{m}}}\ket{\boldsymbol{l}},
\end{align}
which simplifies to
\begin{align}
 \Tr(\frac{d\boldsymbol{ \brho}}{dt}\boldsymbol{ \brho}^{-1}) = \sum_{m=1}^{d}\frac{dp_m}{dt}p_m^{-1}  + \sum_{m=1}^{d} \bra{\boldsymbol{m}}\ket{d_t{\boldsymbol{m}}} +  \sum_{m=1}^{d} \bra{d_t{\boldsymbol{m}}}\ket{\boldsymbol{m}} = \sum_{m=1}^{d}\frac{d\ln(p_m)}{dt} 
\end{align}
using $\bra{d_t{\boldsymbol{m}}}\ket{\boldsymbol{m}} + \bra{\boldsymbol{m}}\ket{d_t{\boldsymbol{m}}}  = d_t{\bra{\boldsymbol{m}}\ket{\boldsymbol{m}}} = 0$.
Recognizing
\begin{align}
 \Tr(\frac{d\boldsymbol{ \brho}}{dt}\boldsymbol{ \brho}^{-1})   =  \frac{d}{dt}\ln(\prod_{m=1}^{d}p_m)  = \frac{d\ln(\lvert\brho\rvert)}{dt}=\frac{d\lvert\brho\rvert}{dt}\lvert \brho\rvert^{-1} 
\end{align}
we can rearrange to find Jacobi's formula for the mixed state
\begin{align}
    \frac{d\lvert\brho\rvert}{dt} =  \lvert \brho\rvert\Tr( \brho^{-1}\frac{d\brho}{dt}).
\end{align}
%\jrgc{Switched from total to partial derivatives here? \msc{Fixed}}
The mixed state $\brho$ must satisfy the Lyapunov equation 
\begin{align}
   \frac{d\brho}{dt} = \frac{\boldsymbol{L}\brho +  \brho\boldsymbol{L}^\top}{2},
\end{align}
where $\boldsymbol{L}$ is the logarithmic derivative. Thus, taking the trace of both sides of
\begin{align}
\brho^{-1}\left(d_t\brho\right) = \frac{ \brho^{-1}\boldsymbol{L}\brho +   \brho^{-1}\brho\boldsymbol{L}^\top}{2}
    = \frac{ \brho^{-1}\boldsymbol{L}\brho +  \boldsymbol{L}^\top}{2},
\end{align}
we get
\begin{align}
 \Tr\left(\brho^{-1}\left(d_t\brho\right)\right) = \Tr\left(\boldsymbol{L}\right),  
\end{align} 
%\jrgc{Doesn't this assume the log derivative is symmetric?} \msc{ I don't think so because $\Tr\left(\brho^{-1}\boldsymbol{L}\brho\right) =\Tr\left(\boldsymbol{L}\right)=\Tr\left(\boldsymbol{L}^\top\right)$}
which we can relate Jacobi's formula in Eq.~\ref{eq:jcb1},
\begin{align}\label{eq:jcb2}
   \frac{d\lvert\brho\rvert}{dt} = \lvert\brho\rvert \Tr\left(\boldsymbol{L}\right) ,
\end{align}
and Eq.~\ref{eq:jc15}.
 
%\section{To show:~\texorpdfstring{$d_t \Tr \bxi = 2 \Tr(\stability_+\bxi)$}{i}}\label{SM:trace-der-relative}

%\jrgc{I think we discussed cutting this Appendix and citing Swet's 2022 PRE for the proof.}
 
%The equation of motion for an unnormalized density matrix $\bxi_i$ is given by (Ref.~\cite{DasGreen2022}):
%\begin{align}
 %\frac{d}{dt}\bxi_i = & = \{\stability_+,\bxi\} + [\stability_-,\bxi].
%\end{align}
%The time derivative of the mixed density matrix $\bxi = \sum_{i=1}^N d_t\bxi_i $ is then given by:
%\begin{align}
 %\frac{d}{dt}\bxi &= \sum_{i=1}^N \frac{d}{dt}\bxi_i =  \sum_{i=1}^N \left(\{\stability_+,\bxi_i\} + [\stability_-,\bxi_i]\right) \nonumber\\
 %& = \{\stability_+,\bxi\} + [\stability_-,\bxi].
%\end{align}
%Hence, the time derivative of $\Tr \bxi$ is given by:
%\begin{align}
%\frac{d}{dt}\Tr\bxi = 2\Tr(\stability_+\bxi).
%\end{align}
%\noindent Dividing both sides by $\Tr \bxi$ yields: 
%\begin{align}
%\frac{1}{\Tr\bxi}\frac{d}{dt}\Tr\bxi = \frac{2}{\Tr\bxi}\Tr(\stability_+\bxi) = 2\Tr(\stability_+\frac{\bxi}{\Tr\bxi}).
%\end{align}
%Therefore, we get:
%\begin{align}
%\frac{d}{dt}\ln \Tr \bxi = 2\Tr(\stability_+\brho).
%\end{align}

\section{Gibbs entropy rate for classical mixed states}\label{SM:E13}

To derive this Gibbs (differential Shannon) entropy rate, we start from its definition
\begin{align}
	S_G =-k_B\int \rho(\ex)\ln\rho(\ex)\,{d\ex} = -k_B\left\langle\ln\rho\right\rangle_{\rho}
\end{align} 
with averages $\langle\cdot\rangle_{\rho}$. Taking this distribution $\rho$ to be the multivariate normal distribution $\rho_G$ (with $k$ variables), the differential entropy is
\begin{equation}
\begin{aligned}
	\mathcal{S}_G &= -k_B \left\langle\ln\left(\frac{\text{exp}(-\frac{1}{2}(\ex - \ex_0)^\top \bm\chi^{-1}(\ex - \ex_0))}{\sqrt{(2\pi)^k|\bm\chi|}}\right) \right\rangle_{\rho_G}\\
    &= \frac{k_B}{2}\ln\left(\left(2\pi\right)^k|\bm\chi|\right) + \frac{k_B}{2}\langle \left(\ex - \ex_0\right)^\top \bm\chi^{-1}\left(\ex - \ex_0\right) \rangle_{\rho_G}\\
    &=  \frac{k_B}{2}\ln\left(\left(2\pi\right)^k|\bm\chi|\right) + \frac{k_B}{2}\langle \Tr\left(\bm\chi^{-1}\left(\ex - \ex_0\right)\left(\ex - \ex_0\right)^\top\right)\rangle_{\rho_G}\\
    &=  \frac{k_B}{2}\ln\left(\left(2\pi\right)^k|\bm\chi|\right) + \frac{k_B}{2} \Tr\left(\bm\chi^{-1}\int\rho_G(\ex)\left(\ex - \ex_0\right)\left(\ex - \ex_0\right)^\top\,{d\ex}\right) \\
    &= \frac{k_B}{2}\ln\left(\left(2\pi\right)^k|\bm\chi|\right) + \frac{k_B}{2} \Tr\left(\bm\chi^{-1}\bm\chi\right) \\
	&=  \frac{k_B}{2}\ln\left(\left(2\pi\right)^k|\bm\chi|\right) + \frac{k_B}{2} \Tr\left(\boldsymbol{\mathbb{1}}_k\right) \\
    &=  \frac{k_B}{2}\ln\left(\left(2\pi\right)^k|\bm\chi|\right) + \frac{k_Bk}{2}\\
    &=  \frac{k_B}{2}\ln\left(\left(2\pi{e}\right)^k|\bm\chi|\right)       
\end{aligned}
\end{equation}
We use the typeface $\mathcal{S}_G$ to indicate this specific form of the Gibbs entropy $S_G$.
Because the density matrix is proportional to the covariance matrix $\boldsymbol{\chi}=c\brho$, the two matrices have the same dimensionality $k=d$ and $|c\brho| = c^d |\brho|$. The Gibbs entropy becomes
\begin{equation}
    \mathcal{S}_G =  \frac{k_B}{2}\ln\left(\left(2\pi{e}c\right)^d|\brho|\right) = \frac{k_B}{2}\ln\left(\left(2\pi{e}c\right)^d\right) + \frac{k_B}{2}\ln|\brho|.
\end{equation}
Since the proportionality constant $c$ is time independent, the entropy rate
%\begin{equation}
%     \dot{\mathcal{S}}_G/k_B =    \frac{1}{2}\frac{d}{dt} \ln |\brho|
%\end{equation}
is directly proportional the contraction rate in Jacobi's formula for the mixed state, and we get
\begin{equation}
     \dot{\mathcal{S}}_G/k_B = \frac{1}{2}\frac{d}{dt} \ln |\brho|,
\end{equation}
Eq.~\ref{eq:gibbs2}.
 
\end{document}